\documentclass[12pt]{article}

\usepackage{geometry}
\usepackage{sectsty}
\usepackage{authblk}
\usepackage{amssymb}
\usepackage{braket}
\usepackage{amsmath}
\usepackage{simplewick}
\usepackage{bm}
\usepackage{graphicx}

\allsectionsfont{\sffamily}

\def\hh{\hspace{0.5mm}}
\def\hs{\hspace{5mm}}
\def\dd{\mbox{d}}
\def\ddd{\text{\scriptsize d}}
\title{\sffamily {\bf Reaction Diffusion Systems and Extensions of Quantum Stochastic Processes}}
\author{Chris D Greenman}
\affil{\small{School of Computing Sciences, University of East Anglia, Norwich,
NR4 7TJ, United Kingdom. \\ email: C.Greenman@uea.ac.uk}}
	
\begin{document}

\maketitle

\begin{abstract}
Reaction diffusion systems describe the behaviour of dynamic, interacting, particulate systems. Quantum stochastic processes generalise Brownian motion and Poisson processes, having operator valued It\^{o} calculus machinery. Here it is shown that the three standard noises of quantum stochastic processes can be extended to model reaction diffusion systems, the methods being exemplified with spatial birth-death processes. The usual approach for these systems are master equations, or Doi-Peliti path integration techniques. The machinery described here provide efficient analyses for many systems of interest, and offer an alternative set of tools to investigate such problems.
\end{abstract}

%%%%%%%%%%%%%%%%%%%%%%%%%%%%%%%%%%%%%%%%%%%%%%%%%%%%%%%%%%%%%

\section{Introduction}

Reaction diffusion systems involve models with discrete entities that have continuous modes of location, transport and interaction \cite{Cardy2006, Liggett1985, Tauber2014}. The classic cases are those of molecular reactions, where molecules diffuse through a medium, and are able to interact in some fashion, once reactive molecules move within sufficient proximity, for example \cite{Doi1976b, Kirkwood1946, Kirkwood1947}. Other examples include birth-death processes where members of the population have spatial dependencies \cite{Dobramysl2018}, and age-structured populations \cite{Greenman2016, Greenman2017, Greenman2015}. 

The methods of analysis for these systems are varied. Classical master equations can be constructed for these systems. This usually results in an infinite series of equations to solve, known as BBGKY hierarchies \cite{Bogoliubov1946, Born1949, Kirkwood1946, Kirkwood1947, Yvon1935}, which are difficult to solve, and so approximation techniques such as moment closure are applicable \cite{Kuehn2016}. Doi developed continuous field theory techniques adapting the machinery from quantum field theory \cite{Doi1976a,Doi1976b}. Notably, Peliti \cite{Peliti1985} developed a lattice framework (rather than Doi's continuous approach) to construct path integral techniques. Together these have seen a wealth of applications, including glass phase transition \cite{Garrahan2007}, branching random walks \cite{Cardy1998}, phylogeny \cite{Jarvis2005}, age structured systems \cite{Greenman2016, Greenman2017, Greenman2015}, neural network fluctuations \cite{Buice2007}, exclusion processes \cite{Greenman2018, Schulz2005, Schulz1996}, predator prey systems \cite{Dobramysl2018}, stochastic duality \cite{Greenman2020, Liggett1985, Ohkubo2013}, knot diagram dynamics \cite{Rohwer2015}, to name a few. See  \cite{Tauber2014} for a broad coverage of related topics. Other approaches that have seen utility in this area include configuration spaces \cite{Lenard1975, Lenard1975b}, system size expansion methods \cite{VanKampen1992} and techniques from semi-groups \cite{Engel2000}. Whatever the technique, the variable dimension size makes the analyses difficult and finding alternative techniques is always a worthwhile endeavour. 

Quantum stochastic processes can be viewed in a generalisation of Brownian motion and Poisson processes. Brownian motion in particular has an It\^{o} calculus associated with it that has seen a plethora of applications across physics, finance and biology \cite{Klebaner2012}.  Quantum stochastic processes also have an associated It\^{o} calculus machinery, although the processes involved are operator rather than vector valued. These techniques most commonly find application in quantum optics \cite{Dalibard1992} and open quantum systems with heat baths \cite{Gardiner1985, Waldenfels2013}. Here we report a correspondence between Doi field theory and quantum stochastic processes. This has the advantage that quantum It\^{o} calculus can be brought to bear on reaction diffusion systems. For local processes (where interactions occur when entities meet at a single location), this machinery allows probability generating functions to be constructed exactly without recourse to field theory methods such as path integrals. For non-local processes (where interactions occur between physically separated entities), perturbation expansions are still necessary (as they are in field theory methods). However, the expansions can also be formulated with the aid of It\^{o} calculus. This work thus reveals a new set of tools  that reaction diffusion systems can be approached with.

The operator valued nature of quantum stochastic processes means extracting probabilities involves an extra step (an inner product). This is also a fundamental feature of obtaining probabilities in quantum mechanical systems, random matrix theory, as well as Doi-Peliti methods. The study of such properties in systems is generally known as non-commutative probability (or sometimes quantum probability) \cite{Chang2015, Meyer1995}. It has also seen development in abstract settings such as graph theory, where it is also referred to as algebraic probability \cite{Hora2007}. 

Now, multi-particle systems are described by Fock spaces, which can be discrete (when particle numbers are modelled) or continuous (where spatial properties are incorporated) in nature. The simplest discrete Fock space is perhaps formed from the ladder operators seen in quantum harmonic oscillators. In \cite{Hora2007} these are extended, where Jacobi sequences are employed to give more general ladder processes applied to graph spectral theory. Discrete Doi-Peliti methods \cite{Peliti1985} also employ such structures to model birth-death processes. These are also extended in \cite{Ohkubo2013b}, meaning a broader class of discrete random processes can be modelled (including fermionic systems that model exclusive processes). Continuous Fock spaces are described by quantum stochastic processes \cite{Chang2015, Eyre2006, Holevo2003, Meyer1995, Milz2021, Parthasarathy1992}. As well as applications to heat baths and optics, connections with classical stochastic processes are also known, where natural formulations of Brownian motion and Poisson processes can be constructed. Furthermore, an Evans-Hudson flow can be formed to describe birth-death processes \cite{Fagnola1991, Parthasarathy1990}, although these do not exhibit the spatial dependency seen in reaction diffusion systems. 

The spacial aspect of continuous Fock spaces thus opens the possibility of utilising quantum stochastic process techniques for reaction diffusion systems. There are three standard noise processes in quantum stochastic processes;  the creation, annihilation and gauge (or number) operators \cite{Coquio2000}. We show that the vast range of classical operators used in Doi field theory can be interpreted as quantum stochastic noise processes, along with corresponding modes of multiplication (that is, an associated It\^{o} table). This involves the development of additional quantum noise processes, although some standard behaviour can be lost. In particular, multiplication in the corresponding It\^{o} tables are not always associative. However, these constructions still provide utility, and lead to alternative methods of analysis for some reaction diffusion systems that usually entail path integrals.

The next two sections give brief overviews of Doi field theory used in reaction diffusion systems, and the Fock spaces analysed in quantum stochastic processes. A mapping between quantum stochastic processes and Doi field theory is then described. Applications to a range of local birth-death and reaction diffusion processes then follow. Perturbation theory for non-local processes are then considered, and conclusions complete the work. 

%%%%%%%%%%%%%%%%%%%%%%%%%%%%%%%%%%%%%%%%%%%%%%%%%%%%%%%%%%%%%
%%%%%%%%%%%%%%%%%%%%%%%%%%%%%%%%%%%%%%%%%%%%%%%%%%%%%%%%%%%%%

\section{Field Theory}

In all that follows, a population of mobile interacting discrete entities is considered. These are referred to as \emph{particles}, although they could be molecules, individuals or any other class of objects modelled this way. They are assumed to be identical and indistinguishable. The particles are also presumed to exist in some space $\Upsilon$. For example, this could be $\mathbb{R}^3$ for spatial variation, $\mathbb{R}^+$ for age dependency, or more general spaces of interest. This property is referred to as the particle's \emph{position}.

There are two general approaches to field theory for these systems. One is the original continuous methods used by Doi \cite{Doi1976a, Doi1976b}, the other involves lattice representation as described by Peliti \cite{Peliti1985}. We consider both representations below. However, although the latter method has probably seen more applications, we will see that the continuous representation of Doi is more naturally interpretable as a quantum stochastic processes and will be the main representation used.

%%%%%%%%%%%%%%%%%%%%%%%%%%%%%%%%%%%%%%%%%%%%%%%%%%%%%%%%%%%%%

\subsection{Doi Continuous Field Theory} 

Doi field theory will now be used to portray a population of interacting particles moving in a spatial continuum. This representation starts with the \emph{vacuum} state, representing an empty population. This is denoted by the \emph{bra} $\bra{\phi}$ or dual \emph{ket} state $\ket{\phi}$, which are given normalisation,
\begin{equation}
\braket{\phi|\phi} = 1.
\label{FundNorm}
\end{equation}

To describe particles and interactions, we introduce (bosonic) creation $a_q^\dag$ and annihilation $a_p$ operators, which act on the vacuum state via,
\begin{equation}
\bra{\phi}a_q^\dag = 0,
\hs
a_p\ket{\phi} = 0,
\nonumber
\end{equation}
and satisfy commutation relations,
\begin{equation}
[a_p,a_q^\dag] = \delta(p-q),
\hs
[a_p,a_q] = 0,
\hs
[a_p^\dag,a_q^\dag] = 0.
\label{Comm}
\end{equation}
Although $a_p$ kills the vacuum state $\ket{\phi}$, the creation operators do not, and so we find states of the following form, 
\begin{eqnarray}
\ket{{\bf p}_n} & \equiv & \ket{p_1,p_2,\dots,p_n}
\equiv a_{p_1}^\dag a_{p_2}^\dag\dots a_{p_n}^\dag\ket{\phi},
\nonumber\\
\bra{{\bf p}_n} & \equiv & \bra{p_1,p_2,\dots,p_n}
\equiv \bra{\phi}a_{p_1}a_{p_2}\dots a_{p_n}.
\nonumber
\end{eqnarray}
Notionally, the creation operators are creating particles, and the state $\ket{{\bf p}_n}$ represents a population of $n$ indistinguishable individuals with locations $p_1,p_2,\dots,p_n$. The commutation relations and normalisation in Eq. (\ref{FundNorm}) can then be used to reveal single particle normalisation $\braket{p|q} = \delta(p-q)$, and more generally (where $S_n$ denotes the symmetric group),
\begin{equation}
\braket{{\bf q}_m|{\bf p}_n} = \delta_{m,n}\sum_{\sigma \in S_n}\prod_{i=1}^n\delta(q_i-p_{\sigma(i)}).
\label{KetNorm}
\end{equation}

This machinery provides a description for a known, fixed configuration of $n$ particles. To model a stochastic system, a time dependent state is defined as,
\begin{equation}
\ket{\psi_t} = \sum_{n=0}^\infty \iint_{\Upsilon^n}\frac{\dd {\bf p}_n}{n!} f^{(n)}({\bf p}_n;t)\ket{{\bf p}_n},
\nonumber
\end{equation}
where the symmetric function $f^{(n)}({\bf p}_n;t)\hh \dd {\bf p}_n$ is the probability density for the population containing $n$ particles at time $t$, such that a labelling $1,2,\dots,n$ exists where the $i^\mathrm{th}$ particle lies in interval $[p_i,p_i+\dd p_i]$ (by `interval', this is the hypercube from $p_i$ to $p_i+\dd p_i$ in $\Upsilon$; see \cite{Doi1976a, Greenman2017} for a more detailed description). The density can be extracted with the aid of Eq. (\ref{KetNorm}) to find,
\begin{equation}
f^{(n)}({\bf p}_n;t) = \braket{{\bf p}_n|\psi_t}.
\label{DensEx}
\end{equation}

Now, the dynamics are described by the evolution equation,
\begin{equation}
\ket{\psi_t} = e^{t\mathcal{L}}\ket{\psi_0},
\nonumber
\end{equation}
for a suitable initial distribution $\ket{\psi_0}$ and evolution operator $\mathcal{L}$. This is often referred to as a Liouvillian operator, and the form taken by $\mathcal{L}$ incorporates both interactions that can change the number and class of particles, and also the dynamics of the mobile particles in space $\Upsilon$. 

For example,
\begin{equation}
\mathcal{L} = \int_\Upsilon \left(\mu(p)\left((a_p^\dag)^2a_p -a_p^\dag a_p\right) + D(p)a_p^\dag \Delta a_p\right) \dd p, 
\label{BirthDiffLiouv}
\end{equation}
would represent a birth process at rate $\mu(p)$ for particles diffusing at rate $D(p)$. The first term kills a particle and replaces it with two new ones (birth of a single new particle). The second term provides the neutral balancing term in the master equation (note the creation and destruction of a particle having no net effect). The third term describes the dynamics, with Laplacian operator $\Delta$ modelling diffusion.

Some care is needed when incorporating operators into Liouvillians (such as with the Laplacian above). In \cite{Doi1976a} such operators are written $\mathcal{L} = \int_\Upsilon \dd p \hh a_p^\dag G(p) a_p$, where the action of the operator is meant in the sense that (shorthand notation such as $G_p\equiv G(p)$ will frequently be adopted for brevity),
\begin{equation}
\int_\Upsilon \dd p \hh a_p^\dag G_p a_p \cdot \sum_{n=0}^\infty \iint_{\Upsilon^n}\hh\frac{\dd {\bf p}_n}{n!}f^{(n)}({\bf p}_n;t)\ket{{\bf p}_n}
= \sum_{n=0}^\infty \iint_{\Upsilon^n}\hh\frac{\dd {\bf p}_n}{n!}\sum_{i=1}^n\left(G_{p_i}f^{(n)}\right)({\bf p}_n;t)\ket{{\bf p}_n}.
\nonumber
\end{equation}
Thus the position of $G$ between $a_p^\dag$ and $a_p$ is not important, with $a_p^\dag a_p$ acting on the ket and $G$ acting on the coefficient (see \cite{Doi1976a} for a more general description).
 
Lastly, coherent states are introduced. The coherent state $\ket{v}$, for an integrable function $v$ acting on $\Upsilon$, is defined as,
\begin{equation}
\ket{v} = e^{\int_\Upsilon v(p)a_p^\dag \hh \ddd p}\ket{\phi}
= \sum_{n=0}^\infty \iint_{\Upsilon^n}\frac{\dd {\bf p}_n}{n!} v(p_1)\dots v(p_n) \ket{{\bf p}_n}.
\label{CohDef}
\end{equation}
The coherent states have the following eigenfunction and normalisation properties,
\begin{eqnarray}
a_p\ket{v} & = & v(p)\ket{v},
\nonumber \\
a_p^\dag \ket{v} & = & \frac{\delta}{\delta v(p)}\ket{v},
\nonumber \\
\braket{u|v} & = & \exp\left\{\int_\Upsilon \bar{u} v \hh \dd p\right\}.
\label{CohNorm}
\end{eqnarray}
Note that the second equation is a functional derivative meant in the sense that $\braket{u|a_p^\dag|v} =  \frac{\delta}{\delta v(p)}\braket{u|v}$. The third equation tells us that for  $v\in L_2(\Upsilon)$, the state $\ket{v}$ is normalisable. In all that follows, coherent state functions (such as $u$ from $\bra{u}$ above) shall be assumed real and conjugation can safely be ignored.

One coherent state of particular importance (that is not generally normalisable) is the coherent state $\ket{1}$ for the constant function $1$. This plays a crucial role in finding moments. For example, probability conservation, and the $m$ point correlation function are given by,
\begin{eqnarray}
1 & = & \braket{1|\psi_t},
\label{TotProb} \\
X_m({\bf p}_m;t) & = & \braket{1|a_{p_1}\dots a_{p_n}|\psi_t},
\label{DoiMom}
\end{eqnarray}
where $X_m({\bf p}_m;t)\dd {\bf p}_m$ is the probability we can find and label $m$ particles (from a possibly larger population) such that the $i^\mathrm{th}$ lies in the interval $[p_i,p_i+\dd p_i]$. The function $X_1(p;t)$ is thus the regular density. Note also from Eq.s (\ref{CohNorm}) and (\ref{TotProb}) that for a coherent state to be probability state, it must be normalised properly. For an initial probability state we could use, for example,
\begin{equation}
\ket{\psi_0} = e^{-\int_\Upsilon v(p) \hh \ddd p}\ket{v}.
\label{Init}
\end{equation}

Further properties of the Doi framework can be found in \cite{Doi1976a, Greenman2017}.

%%%%%%%%%%%%%%%%%%%%%%%%%%%%%%%%%%%%%%%%%%%%%%%%%%%%%%%%%%%%%

\subsection{Peliti Lattice Methods}
\label{Lattice}

For lattice methods, the changes are not substantial and we just point out the main differences. The space $\Upsilon$ is now assumed to be a discrete regular cubical lattice that particles hop about and interact upon. The distance between nearest neighbour sites of the lattice is known as the \emph{lattice constant} $\epsilon$, which is usually later reduced to zero in a limiting process to the continuum. The lattice is characterised by a set of integer \emph{occupation numbers} $\{n_i\}$ counting the number of particles occupying site $i$. To represent movements and interactions, we now use a bosonic ladder operator algebra, with annihilation and creation operators $a_i$ and $a_i^\dag$ that satisfy commutation relations, 
\begin{equation}
[a_i,a_j^\dag] = \delta_{ij},
\hs
[a_j,a_j] = 0,
\hs
[a_i^\dag,a_j^\dag] = 0.
\nonumber
\end{equation}

Fundamental states of the system are represented as $\ket{\{n_i\}} \equiv \ket{\dots,n_i,\dots}$, where $n_i$ counts the number of particles occupying site $i$, and the operators have actions,
\begin{eqnarray}
a_j\ket{\dots,n_i,\dots} & = & n_j\ket{\dots,n_i-1,\dots},
\nonumber\\
a_j^\dag\ket{\dots,n_i,\dots} & = & \ket{\dots,n_i+1,\dots}.
\nonumber
\end{eqnarray}

Coherent states are very similar to the continuous formalism, and can be defined as,
\begin{equation}
\ket{u} = \prod_{i \in \Upsilon}e^{u_ia_i^\dag}\ket{\phi},
\nonumber
\end{equation}
which is just a discretised form of Eq. (\ref{CohDef}), from which standard eigenfunction properties such as $a_i\ket{u} = u_i\ket{u}$ follow.

The general state of a stochastic system is then represented as,
\begin{equation}
\ket{\psi_t} = \sum_{\{n_i\}}P(\{n_i\};t)\ket{\{n_i\}},
\nonumber
\end{equation}
where $P(\{n_i\};t)$ represents the probability that state occupation numbers are $\{n_i\}$ at time $t$. The dynamics  are similar to the continuous formalism, where we have the form,
\begin{equation}
\frac{\partial \ket{\psi_t}}{\partial t} = \mathcal{L}\ket{\psi_t},
\nonumber
\end{equation}
with Liouvillian $\mathcal{L}$ capturing movements and interactions.

For example, in $\mathcal{L} = \lambda\sum_i (1-a_i^{\dag^2})a_i^2+D\sum_{<i,j>}(a_i^\dag-a_i^\dag)(a_i-a_j)$, the first term would represent pairwise annihilation $A+A \rightarrow 0$ (for pairs of particles annihilating on the same site) and the second term represents a hoping process to nearest neighbours ($<i,j>$ represents nearest neighbour pairs). In the limit to the continuum this term becomes a diffusion term, although some care is needed in the limiting process $\epsilon \rightarrow 0$ as we need to rescale coefficients correctly. For example, $D\epsilon^2$ would become the diffusion rate coefficient (see \cite{Cardy2006, Peliti1985, Tauber2014} for more details). 

%%%%%%%%%%%%%%%%%%%%%%%%%%%%%%%%%%%%%%%%%%%%%%%%%%%%%%%%%%%%%
%%%%%%%%%%%%%%%%%%%%%%%%%%%%%%%%%%%%%%%%%%%%%%%%%%%%%%%%%%%%%

\section{Fock Space}

Fock space provides a framework that more rigorously describes the space used in the previous section. This section gives a brief overview of the important features. Note that it follows the framework used in Meyer \cite{Meyer1995}, the construction used in Parthasarathy \cite{Parthasarathy1992} differs slightly. There is a basic Hilbert space $\mathcal{H}$. Elements of this space represent the single particle configurations (that is, elements of $\mathcal{H}$ act on $\Upsilon$). For $n$ particle configurations, we have elements in the tensor product $\mathcal{H} \otimes \mathcal{H} \otimes \dots \otimes \mathcal{H} = \mathcal{H}^{\otimes n}$. This is used to describe configurations where particles are labelled (that is, individual particles can be identified and differentiated).  

For the case where particles cannot be differentiated, we consider the subspace of $\mathcal{H}^{\otimes n}$ formed by symmetric components. Specifically, terms of the following form,
\begin{equation}
u_1\circ u_2 \circ \dots \circ u_n = \frac{1}{n!}\sum_{\sigma \in S_n}u_{\sigma(1)}\otimes
u_{\sigma(2)}\otimes \dots \otimes u_{\sigma(n)}. 
\nonumber
\end{equation}
This space is denoted $\mathcal{H}^{\circ n}$ and represents $n$ \emph{particle space}, also known as the $n^\mathrm{th}$ \emph{symmetric power} of $\mathcal{H}$, or more archaically as the $n^\mathrm{th}$ \emph{chaos}. This is the space used to represent bosonic systems in quantum mechanics and is the form used throughout this work. There also exists antisymmetric space, used to represent fermionic systems \cite{Meyer1995, Parthasarathy1992}, although these will not play any role in this work. We also have the $0^\mathrm{th}$ symmetric space defined as $\mathcal{H}^{\circ 0} = \mathcal{H}_0 = \mathbb{C}$. The element $1 \in \mathcal{H}_0$ is known as the \emph{vacuum vector}.

The Fock space is then formed by taking the direct sum of these spaces, denoted as,
\begin{equation}
\Gamma(\mathcal{H}) = \oplus_{n=0}^\infty \mathcal{H}^{\circ n}.
\nonumber
\end{equation}

Now the space $\mathcal{H}$ is endowed with an inner product, denoted $<\cdot,\cdot>_\mathcal{H}$, with subscript to distinguish from bras and kets. This has a natural extension to $n$ particle space, given by the permanent of individual inner products,
\begin{equation}
<u_1 \circ u_2 \circ \dots \circ u_n, v_1 \circ v_2 \circ \dots \circ v_n>_{\circ n} 
= \mathrm{per}(<u_i,v_j>_\mathcal{H})
= \sum_{\sigma\in S_n}\prod_{i=1}^n<u_i,v_{\sigma(i)}>_\mathcal{H}.
\nonumber
\end{equation}
For example, if $f^{\circ n}$ and $g^{\circ n}$ represent the symmetric product of $n$ copies of the functions $f,g \in \mathcal{H}$, then,
\begin{equation}
<f^{\circ n},g^{\circ n}>_{\circ n} = n!(<f,g>_\mathcal{H})^n.
\label{SymNormRes}
\end{equation}

Then for the full Fock space, taking elements $F = \oplus_{n=0}^\infty f^{(n)}$ and $G = \oplus_{n=0}^\infty g^{(n)}$, where $f^{(n)},g^{(n)}\in \mathcal{H}^{\circ n}$, we define inner product,
\begin{equation}
<F,G>_\Gamma = \sum_{n=0}^\infty <f^{(n)},g^{(n)}>_{\circ n}.
\label{GammaNorm}
\end{equation}
If we abuse notation and also write $f^{(n)}\in \mathcal{H}^{\circ n}$ for Fock space element $0\oplus \dots \oplus 0 \oplus f^{(n)} \oplus 0 \oplus \dots \oplus 0 \in \Gamma(\mathcal{H})$, we find $<f^{(m)}, g^{(n)}>_\Gamma = \delta_{m,n}<f^{(m)}, g^{(m)}>_{\circ m}$. That is, firstly, elements from distinct symmetric spaces are orthogonal, and secondly, that the inner product on $\Gamma(\mathcal{H})$ is the naturally induced one from the symmetric spaces.
  
Now, \emph{exponential vectors} are a subset of vectors in Fock space that are well behaved. They are defined as,
\begin{equation}
\xi(u) = \sum_{n=0}^\infty\frac{u^{\circ n}}{n!}.
\label{ExpDef}
\end{equation}
Note in particular from Eq.s (\ref{SymNormRes}) and (\ref{GammaNorm}) that we have normalisation,
\begin{equation}
<\xi(u),\xi(v)>_\Lambda = \exp<u,v>_\mathcal{H}.
\nonumber
\end{equation}
The space generated by exponential vectors is a dense subspace of $\Gamma(\mathcal{H})$ and subsequently proves to be the testing ground for many results in Fock space \cite{Meyer1995, Parthasarathy1992}, but they also behave well under the action of important operators. 

In particular, we have annihilation, creation and gauge operators that act as follows,
\begin{eqnarray}
A_f(u_1 \circ \dots \circ u_n) & = & \sum_{i=1}^n<f,u_i>_\mathcal{H}u_1 \circ \cdots \hat{u}_i \dots \circ u_n,
\nonumber \\
A_g^\dag(u_1 \circ \dots \circ u_n) & = & g \circ u_1 \circ \dots \circ u_n,
\nonumber \\
\Lambda_H(u_1 \circ \dots \circ u_n) & = & H(u_1) \circ u_2 \circ \dots \circ u_n + u_1 \circ H(u_2) \circ \dots \circ u_n + \dots + u_1 \circ u_2 \circ \dots \circ H(u_n),
\nonumber
\end{eqnarray}
where $\hat{u}_i$ indicates an element removed from the product, $H$ is a linear operator acting on $\mathcal{H}$, and $f,g \in \mathcal{H}$. These operators have well behaved actions on exponential vectors. Specifically,
\begin{eqnarray}
A_f\xi(u) & = & <f,u>_\mathcal{H}\xi(u),
\nonumber \\
A_g^\dag \xi(u) & = & \lim_{\epsilon\to 0}\frac{\partial}{\partial \epsilon}\xi(u+\epsilon g),
\nonumber \\
\Lambda_H \xi(u) & = & A_{Hu}^\dag \xi(u).
\label{ActsFock}
\end{eqnarray}

Now to construct quantum stochastic processes with associated noises, the Fock space usually employed is $\Gamma(L_2(\mathbb{R}^+))$. Functions taken from space $\mathcal{H}=L_2(\mathbb{R}^+)$ are then one dimensional square integrable functions that are usually viewed as acting on time $t$. The It\^{o} calculus associated with Brownian motion can now be generalised to this space. This first requires the restriction offered by indicator functions, where,
\begin{equation}
\chi_{[a,b]}(t)=
\begin{cases}
1, & \text{if $t \in [a,b]$}.\\
0, & \text{otherwise}.
\end{cases}
\nonumber
\end{equation}
Then $f\chi_{[0,t]}$ and $f\chi_{[t,t+\dd t]}$ represent the function $f$ restricted to support from intervals $[0,t]$ and infinitesimal $[t,t+\dd t]$, respectively. Associated with these, the time dependent operator $A_f(t) = A_{f\chi_{[0,t]}}$ and differential $\dd A_f(t) = A_{f\chi_{[t,t+\ddd t]}}$ are defined, with similar expressions for creation and gauge operators. It is these differentials that follow It\^{o} multiplication rules, as given in Table \ref{ItoTable}.
\begin{table}[h]
\begin{center}
\begin{tabular}{||c||cccc||} 
\hline
& $\dd\Lambda_G$ & $\dd A_g$ & $\dd A_g^\dag$ & $\dd t$ \\ [0.5ex] 
\hline\hline
$\dd\Lambda_F$ & $\dd\Lambda_{FG}$ & $0$ & $\dd A_{Fg}^\dag$ & $0$ \\ 
\hline
$\dd A_f$ & $\dd A_{G^\dag f}$ & $0$ & $<f,g>_\mathcal{H}\dd t$ & $0$ \\
\hline
$\dd A_f^\dag$  & $0$ & $0$ & $0$ & $0$ \\
\hline
$\dd t$ & $0$ & $0$ & $0$ & $0$ \\
\hline
\end{tabular}
\caption{Standard quantum stochastic process It\^{o} table.}
\label{ItoTable}
\end{center}
\end{table}

From this machinery, processes of interest can be constructed. For example, consider the process $\dd X = \dd \Lambda + \sqrt{\ell}(\dd A + \dd A^\dag) + \ell \hh\dd t$ (function and operator subscripts are defined as unity and dropped for a moment). This gives a realisation of the Poisson process (It\^{o} table multiplications readily give $\dd X^2 = \dd X$), and $\dd A + \dd A^\dag$ gives a realisation of Brownian motion (note that $(\dd A + \dd A^\dag)^2=\dd t$). More general birth-death processes have also been constructed \cite{Fagnola1991, Parthasarathy1990}, although (as with many applications) these utilise an extension of Fock space with an \emph{initial} space $\mathcal{H'}$ to $\mathcal{H}'\otimes\Gamma(\mathcal{H})$. We do not require or make use of such extensions.

Next we see how these processes and the Doi framework can be unified, resulting in an alternative method to path integration.

%%%%%%%%%%%%%%%%%%%%%%%%%%%%%%%%%%%%%%%%%%%%%%%%%%%%%%%%%%%%%
%%%%%%%%%%%%%%%%%%%%%%%%%%%%%%%%%%%%%%%%%%%%%%%%%%%%%%%%%%%%%

\section{Non-Associative Ito Calculus and Normal Ordering}

Now, some of the operators seen in Doi calculus directly correspond to standard quantum stochastic noise processes. For this connection, the Fock space $\Gamma(L_2(\Upsilon))$ is used. For quantum stochastic processes, the space $\Upsilon = \mathbb{R}^+$  is usually employed and acts as a time variable. This is not the case here, and $\Upsilon = \mathbb{R}^3$ is the more usual interpretation, representing position, with time a separate variable. Conversely, the initial space $\mathcal{H}'$ that proves so useful in applications plays no role here.

%%%%%%%%%%%%%%%%%%%%%%%%%%%%%%%%%%%%%%%%%%%%%%%%%%%%%%%%%%%%%

\subsection{Doi Field Theory and Quantum Stochastic Process Equivalence}

The correspondence between quantum stochastic processes and Doi field theory can be found from,
\begin{equation}
f_1 \circ \dots \circ f_n \equiv \iint_{\Upsilon^n} \dd {\bf q}_n \hh f(q_1)\dots f(q_n)\ket{{\bf q}_n},
\label{DoiFockCorr}
\end{equation}
where $f_i \in \mathcal{H}$ and $\ket{{\bf q}_n} = a_{q_1}^\dag \dots a_{q_n}^\dag \ket{\phi}$ are the fundamental states in Doi field theory. The vacuum states are also identified under this correspondence. Note that in Fock space, the states $\ket{{\bf q}_n}$ in isolation are singular, and Eq. (\ref{DoiFockCorr}) would require $f_i$ to be delta functions, which do not sit in $\mathcal{H}$. That is, a distribution version of Fock space would be required. Instead, we find Fock states are in effect smeared fundamental Doi states. 

Note from Eq.s (\ref{CohDef}) and (\ref{ExpDef}) that coherent states and exponential vectors are equivalent under the correspondence of Eq. (\ref{DoiFockCorr}). That is,
\begin{equation}
\xi(u) \equiv \ket{u}.
\nonumber
\end{equation}

There is also correspondence between operators. Consider first the annihilation operators, where, 
\begin{equation}
A_f \equiv \int_\Upsilon f(p) a_p \hh\dd p. 
\nonumber
\end{equation}
To see this connection note the equivalent action on exponential vectors and coherent states given by,
\begin{equation}
A_f\xi(u) = <f,u>_\mathcal{H}\xi(u) = \int_\Upsilon f(p) u(p) \hh \dd p \hh \xi(u)
\equiv \int_\Upsilon f(p)u(p) \hh \dd p \ket{u} = \int_\Upsilon f(p)a_p \hh \dd p \ket{u}.
\nonumber
\end{equation}

Secondly, the creation operators have equivalent actions. Specifically, we have operators,
\begin{equation}
A_f^\dag \equiv \int_\Upsilon f(p) a_p^\dag \hh\dd p,
\nonumber
\end{equation}
along with actions,
\begin{equation}
A_f^\dag \xi(u) = \lim_{\epsilon \rightarrow 0}\frac{\partial}{\partial \epsilon}\xi(u+\epsilon f) = \int_\Upsilon f(p)\frac{\delta \xi(u)}{\delta u(p)} \hh \dd p
\equiv
\int_\Upsilon f(p) \frac{\delta}{\delta u(p)} \ket{u}\hh \dd p = \int_\Upsilon f(p) a_p^\dag \hh \dd p \ket{u}.
\nonumber
\end{equation}

Thirdly, we note that the gauge process has the following Doi interpretation,
\begin{equation}
\Lambda_H \equiv \int_\Upsilon a_p^\dag H a_p \hh\dd p,
\nonumber
\end{equation}
with equivalence seen as follows,
\begin{eqnarray}
& & \Lambda_H\xi(u) = A_{Hu}^\dag\xi(u) = \lim_{\epsilon \to 0}\frac{\partial}{\partial \epsilon} \xi(u+\epsilon Hu) = \int_\Upsilon \dd p \hh (Hu)(p)\frac{\delta}{\delta u(p)}\xi(u)
\nonumber \\
& \equiv & \int_\Upsilon \dd p \hh (Hu)(p)\frac{\delta}{\delta u(p)}\ket{u} = 
\int_\Upsilon \dd p \hh (Hu)(p)a_p^\dag\ket{u} = \int_\Upsilon \dd p \hh a_p^\dag H a_p\ket{u}.
\nonumber
\end{eqnarray}

Finally we note that more general operators of Doi field theory do not naturally fit into this formalism. Firstly, the birth operator $\int_\Upsilon \mu(p)(a_p^\dag)^2a_p$ seen in Eq. (\ref{BirthDiffLiouv}) cannot be written in terms of the creation, annihilation or gauge processes of Fock space. Secondly, consider the annihilation process $A+A \rightarrow \phi$, which can be described by Liouvillian $\frac{1}{2}\iint_{\Upsilon^2}R(|p-q|)(a_p^\dag a_q^\dag a_pa_q - a_pa_q) \hh\dd p \hh\dd q$. This describes interaction over a distance, with $R(|p-q|)$ giving the rate of annihilation for particles separated by distance $|p-q|$, and the process is not obviously described by Fock space. Thirdly, different species also occur naturally in such reactions. The conversion process $A \rightarrow B$, which can be described by Liouvillian $\int_{\Upsilon}\mu(p)(b_p^\dag a_p-a_p^\dag a_p) \hh\dd p$ also needs to be adapted to this formalism.

General Fock space methods for these kind of operators are needed. This requires a reinterpretation of Wick ordering, which we now detail.

%%%%%%%%%%%%%%%%%%%%%%%%%%%%%%%%%%%%%%%%%%%%%%%%%%%%%%%%%%%%%

\subsection{Process Multiplication}

To adapt Fock space methods for general operators from Doi field theory, a method to multiply operators is first needed. This can be achieved the aid of Wick's theorem on \emph{normal ordering} of operators, where all creation operators are positioned left of all annihilation operators \cite{Lancaster2014, Maggiore2005, Peskin2018}. Let the parentheses $:\hh:$ indicate normal ordering of operators within, so $:a_pa_q^\dag:=a_q^\dag a_p$, for example. Any annihilation operator $a_p$ left of a creation operator $a_q^\dag$ (that is, a pair of operators not in normal form) are denoted a \emph{contractible pair}, meaning the commutator in Eq. (\ref{Comm}) can be used to turn them into a delta function, and variables $p=q$ can be identified as one variable. Then Wick's theorem tells us that the integral over a product of operators is found by summing the normal ordered product across all choices of contractions. A branch connecting a contractible pair in an equation indicates the pair will be contracted.

For example, consider the operator $a_pa_q^\dag$ that arises when the product $A_fA_g^\dag$ is evaluated. This product is not in normal form, and there are two possible contractions, the null contraction (doing nothing) and the contraction between the pair. Then we write,
\begin{eqnarray}
A_fA_g^\dag & = & \iint_{\Upsilon^2} f(p)g(q)a_pa_q^\dag \hh \dd p \hh \dd q = 
\iint_{\Upsilon^2} f(p)g(q)\left(:a_pa_q^\dag: + 
\contraction[1ex]{:}{a}{_p}{a}
:a_pa_q^\dag:
\right) \hh \dd p \hh \dd q
\nonumber \\
& = & \iint_{\Upsilon^2}f(p)g(q)a_q^\dag a_p \hh \dd p \hh \dd q
+ \int_\Upsilon f(p)g(p) \hh \dd p 
\nonumber.
\end{eqnarray}
Note that the same result can be found by using the commutator in Eq. (\ref{Comm}) directly. 

Now, if $\Upsilon$ is restricted to the infinitesimal hypercube $\dd p$ we find the first term is of order $\dd p^2$ and negligible compared to the second term. This recovers the well known result from quantum stochastic process It\^{o} products (see Table \ref{ItoTable}), $\dd A_f \dd A_g^\dag = f(p)g(p)\dd p$.

Similarly we find the product,
\begin{eqnarray}
\Lambda_G\Lambda_H & = & \iint_{\Upsilon^2} G(p)H(q)a_p^\dag a_p a_q^\dag a_q \hh \dd p \hh \dd q = 
\iint_{\Upsilon^2} G(p)H(q)\left(:a_p^\dag a_p a_q^\dag a_q: + 
\contraction[1ex]{:a_p^\dag}{a}{_p}{a}
:a_p^\dag a_p a_q^\dag a_q:
\right) \hh \dd p \hh \dd q
\nonumber \\
& = & \iint_{\Upsilon^2}G(p)H(q)a_p^\dag a_q^\dag a_p a_q \hh \dd p \hh \dd q
+ \int_\Upsilon G(p)H(p)a_p^\dag a_p \hh \dd p. 
\nonumber
\end{eqnarray}
Again, restricting to infinitesimal $\Upsilon \equiv\dd p$, the first term is negligible and we end up with the familiar product $\dd \Lambda_G \dd \Lambda_H = \dd \Lambda_{GH}$. The other products in standard quantum stochastic It\^{o} tables can be found similarly. 

However, we now find that these have natural extensions for other operators from Doi field theory. For example, consider the operator $\Xi_R = \frac{1}{2}\iint_{\Upsilon^2}R(|p-q|)a_{p}^\dag a_{q}^\dag a_{p}a_{q} \hh \dd p \hh \dd q$ that appeared in the annihilation model $A+A \rightarrow \phi$ above. The product of these operators becomes (using shorthand $R_{p-q}=R(|p-q|)$),
\begin{eqnarray}
\Xi_R\Xi_S & = & \frac{1}{4}\iint_{\Upsilon^4}R_{p-q}S_{p'-q'} \left(
\sum_\mathrm{contractions}:a_{p}^\dag a_{q}^\dag a_{p}a_{q}a_{p'}^\dag a_{q'}^\dag a_{p'}a_{q'}:\right)\hh \dd p \hh \dd q \hh \dd p' \hh \dd q'
\nonumber \\
& = & \frac{1}{4}\iint_{\Upsilon^4}R_{p-q}S_{p'-q'} \left(
:a_{p}^\dag a_{q}^\dag a_{p}a_{q}a_{p'}^\dag a_{q'}^\dag a_{p'}a_{q'}:
+ \contraction[1ex]{:a_{p}^\dag a_{q}^\dag a_{p}}{a}{_q}{a}
:a_{p}^\dag a_{q}^\dag a_{p}a_{q}a_{p'}^\dag a_{q'}^\dag a_{p'}a_{q'}:
\right.
\nonumber \\
& & 
+ \bcontraction[1ex]{:a_{p}^\dag a_{q}^\dag a_{p}}{a}{_{q}a_{p'}^\dag}{a}
:a_{p}^\dag a_{q}^\dag a_{p}a_{q}a_{p'}^\dag a_{q'}^\dag a_{p'}a_{q'}:
+ \contraction[1ex]{:a_{p}^\dag a_{q}^\dag}{a}{_{p}a_{q}}{a}
:a_{p}^\dag a_{q}^\dag a_{p}a_{q}a_{p'}^\dag a_{q'}^\dag a_{p'}a_{q'}:
+\bcontraction[1ex]{:a_{p}^\dag a_{q}^\dag }{a}{_{p}a_{q}a_{p'}^\dag }{a}
:a_{p}^\dag a_{q}^\dag a_{p}a_{q}a_{p'}^\dag a_{q'}^\dag a_{p'}a_{q'}:
\nonumber \\
& &
+ \left.
\contraction[1ex]{:a_{p}^\dag a_{q}^\dag a_{p}}{a}{_q}{a}
\bcontraction[1ex]{:a_{p}^\dag a_{q}^\dag }{a}{_{p}a_{q}a_{p'}^\dag }{a}
:a_{p}^\dag a_{q}^\dag a_{p}a_{q}a_{p'}^\dag a_{q'}^\dag a_{p'}a_{q'}:
+ \contraction[1ex]{:a_{p}^\dag a_{q}^\dag}{a}{_{p}a_{q}}{a}
\bcontraction[1ex]{:a_{p}^\dag a_{q}^\dag a_{p}}{a}{_{q}a_{p'}^\dag}{a}
:a_{p}^\dag a_{q}^\dag a_{p}a_{q}a_{p'}^\dag a_{q'}^\dag a_{p'}a_{q'}:
\right) \hh \dd p \hh \dd q \hh \dd p' \hh \dd q'
\nonumber \\
& = & \frac{1}{4}\iint_{\Upsilon^4}R_{p-q}S_{p'-q'}
a_{p}^\dag a_{q}^\dag a_{p'}^\dag a_{q'}^\dag a_{p}a_{q} a_{p'}a_{q'}
\hh \dd p \hh \dd q \hh \dd p' \hh \dd q'
+\iint_{\Upsilon^3}R_{p-q}S_{q-r}a_{p}^\dag a_{q}^\dag a_{r}^\dag a_p a_q a_r
\hh \dd p \hh \dd q \hh \dd r
\nonumber \\
& & 
+\frac{1}{2}\iint_{\Upsilon^2}R_{p-q}S_{p-q} a_p^\dag a_q^\dag a_p a_q
\hh \dd p \hh \dd q.
\nonumber
\end{eqnarray}
Thus we get seven contractions resulting in three terms. The first term is an integral over $\Upsilon^4$ involving a new operator not of the form $\Xi_R$. The next four terms involve one contraction (or one use of the commutator relation) and result in another new operator, over $\Upsilon^3$. The final two terms have two contractions each (the maximum possible) and result in integrals over $\Upsilon^2$, the same space that each operator $\Xi_R$ and $\Xi_S$ is defined over, resulting in operator $\Xi_{RS}$. Then restricting to infinitesimal $\Upsilon \equiv \dd q$, only the two full contractions survive, the remaining integrals being of order $\dd q^3$ or higher. In particular, we end up with the simple relation,
\begin{equation}
\dd \Xi_R \dd \Xi_S = \dd \Xi_{RS}.
\nonumber
\end{equation}
 
Thus Wick contractions allow us to construct quantum It\^{o} product tables for more general operators seen in Doi field theory. We note that these generalised products are not necessarily associative.

For example, consider the operator $B^{(m)}_G = \int_\Upsilon G(p)(a_p^\dag)^ma_p$. This form of operator can occur in birth processes, as seen in Eq. (\ref{BirthDiffLiouv}) for example. Then multiplying $B^{(m)}_G$ to $B^{(n)}_H$ results in one non-trivial contraction, and we find,
\begin{equation}
B^{(m)}_GB^{(n)}_H = n\int_\Upsilon G(p)H(p)(a_p^\dag)^{m+n-1}a_p \hh \dd p
+ \iint_{\Upsilon^2}G(p)H(q)(a_p^\dag)^m(a_q^\dag)^na_pa_q\hh \dd p \hh \dd q,
\nonumber
\end{equation}
from which we find,
\begin{equation}
\dd B^{(m)}_G \dd B^{(n)}_H = n\hh \dd B^{(m+n-1)}_{GH}.
\label{BProdForm}
\end{equation}
This is not an associative multiplication. In order to put operator products into normal form, the rightmost creation operators are moved left. This has the effect that multiplication of differentials is from the right. For example, one finds that $\dd B^{(\ell)}_F \dd B^{(m)}_G\dd B^{(n)}_H = n(m+n-1)\dd B^{(\ell+m+n-2)}_{FGH}$, as can be shown by explicit expansion into normal form.

The final observation is crucial in the construction of alternative methods to path integration, where we note that \emph{Fock space differentials for distinct positions commute}. If we write, for example, $\dd \Lambda_G(p) = \int_{\Upsilon_p}\dd \Lambda_G$ for the region $\Upsilon_p \equiv \dd p$, then $[\dd \Lambda_G(p), \dd \Lambda_G(q)] = 0$ for $p \ne q$. This is because the delta function in Eq. (\ref{Comm}) cannot be implicated because the regions $\Upsilon_p$ and $\Upsilon_q$ do not overlap. Thus none of the pairwise contractions are possible and only higher order terms remain in the products $\dd \Lambda_G(p) \hh \dd \Lambda_G(q)$ and $\dd \Lambda_G(q) \hh \dd \Lambda_G(p)$, meaning the commutator only contains high order terms and can be regarded as zero.

%%%%%%%%%%%%%%%%%%%%%%%%%%%%%%%%%%%%%%%%%%%%%%%%%%%%%%%%%%%%%

\subsection{Evolution Operator Expansion}

Now, when Doi field theory is employed for many stochastic processes of interest, we have an evolution operator of the form $U=e^{t\mathcal{L}}$, for some Liouvillian operator $\mathcal{L}$. Path integral approaches split time into infinitesimal intervals, and $U$ into a product over small time chunks, with resolutions of identity used to bridge the gaps and form a path integral \cite{Greenman2017, Kleinert2009, Peliti1985}. However, the Fock space calculus described above can instead be used to convert the evolution operator into a useful form. 

Specifically, given that Fock space differentials for distinct loci commute, $U$ can be broken into a product over infinitesimals $\dd p$. That is, if $\mathcal{L} = \int_\Upsilon \dd X$ is written as an integral over Fock space operators, we find,
\begin{equation}
U = e^{t\mathcal{L}} = e^{t\int_\Upsilon \ddd X} = \prod_{\ddd p \in \Upsilon}e^{t\hh\ddd X(p)} 
= \prod_{\ddd p \in \Upsilon}\left(1+\sum_{n=1}^\infty \frac{t^n}{n!}\dd X(p)^n\right).
\nonumber
\end{equation}

For example, consider a simple death-diffusion process $A \rightarrow \phi$ with Liouvillian given by,
\begin{equation}
\mathcal{L} = \int_\chi \left(\mu(p)(a_p-a_p^\dag a_p)+D(p)a_p^\dag\Delta a_p\right) \hh \dd p,
\label{DeathLiouv}
\end{equation}
where the death rate is $\mu(p)$ and particles diffuse at rate $D(p)$. The evolution operator can then be written as follows (where we have operator $H=\mu-D\Delta$),
\begin{eqnarray}
U & = & \exp\left\{ t\int_\Upsilon( \dd A_\mu - \dd\Lambda_\mu + \dd\Lambda_{D\Delta} )\right\} = \exp\left\{ t\int_\Upsilon( \dd A_\mu - \dd \Lambda_H )\right\} 
\nonumber \\
& = & \prod_{\ddd p \in \Upsilon}\exp\left\{ -t( \dd \Lambda_H - \dd A_\mu )\right\} = 
\prod_{\ddd p \in \Upsilon}\left(1+\sum_{n=1}^\infty\frac{(-t)^n}{n!}(\dd \Lambda_H - \dd A_\mu)^n \right).
\nonumber
\end{eqnarray}
The multiplication relations between $\dd A_\mu$ and $\dd \Lambda_H$ are given in Table \ref{ItoTable} and can be used to simplify $(\dd \Lambda_\mu - \dd A_H)^n$. Note that the only surviving terms in an expansion of brackets involve either no $\dd A_\mu$ terms, or a single one on the left. This results in, 
\begin{equation}
U = \prod_{\ddd p \in \Upsilon}\left( 1+\sum_{n=1}^\infty\frac{(-t)^n}{n!}(\dd\Lambda_{H^n} - 
\dd A_{(H^\dag)^{n-1}\mu}) \right)
= \prod_{\ddd p \in \Upsilon}\left( 1+\dd \Lambda_{[e^{-tH}-1]} - \dd A_{[e^{-tH}1-1]} \right).
\label{DeathEvo}
\end{equation}
In the last term we make the simplifying assumption that diffusion $D(p)=D$ is uniform, meaning $H$ is self adjoint and we can put $(H^\dag)^{n-1}\mu = H^n1$, where $H^n$ acts on constant function $1$. The evolution operator is now in a useful form whereby properties such as the moments and probability densities can be extracted.

%%%%%%%%%%%%%%%%%%%%%%%%%%%%%%%%%%%%%%%%%%%%%%%%%%%%%%%%%%%%%

\subsection{Moments}

In order to extract moments from the evolution operator, the initial distribution in Eq. (\ref{Init}) and the $m$ point correlation function of Eq. (\ref{DoiMom}) needs to be interpreted in Fock formalism. 

Firstly, for the initial distribution we have equivalence,
\begin{equation}
\ket{v}e^{-\int_\Upsilon v(p) \hh \ddd p} \equiv \xi(v)e^{-\int_\Upsilon v(p) \hh \ddd p}.
\nonumber
\end{equation}
Note that the normalisation factor requires $v \in L_1(\Upsilon)$, whereas $\|\xi(v)\|_\Gamma<\infty$ requires $v \in L_2(\Upsilon)$. 

The correlation function itself involves singular operators of the form $a_p$, so instead we find that,
\begin{eqnarray}
& & \iint_{\Upsilon^m} \dd {\bf p}_m f_1(p_1)\dots f_m(p_m) X_m({\bf p}_m;t) 
= \iint_{\Upsilon^m} \dd {\bf p}_m f_1(p_1)\dots f_m(p_m)\braket{1|a_{p_1}\dots a_{p_m}|\psi_t}
\nonumber \\
& \equiv & \lim_{u\to 1}<\xi(u),A_{f_1}\dots A_{f_m}U\xi(v)>_\Gamma e^{-\int_\Upsilon v(p) \hh \ddd p}
= \lim_{u\to 1}<A_{f_m}^\dag\dots A_{f_1}^\dag\xi(u),U\xi(v)>_\Gamma e^{-\int_\Upsilon v(p) \hh \ddd p}
\nonumber \\
& = & \lim_{u\to 1,\bm{\epsilon} \to 0}\frac{\partial^m}{\partial \epsilon_1 \dots \partial \epsilon_m}<\xi(u+{\bm{\epsilon}}. {\bf f}),U\xi(v)>_\Gamma e^{-\int_\Upsilon v(p) \hh \ddd p}
\nonumber \\
& = & \left.\iint_{\Upsilon^m}\hh \dd {\bf p}_m \hh f_1(p_1)\dots f_m(p_m)\frac{\delta^m}{\delta u(p_1) \dots \delta u(p_m)}
<\xi(u),U\xi(v)>_\Gamma e^{-\int_\Upsilon v(p) \hh \ddd p}\right|_{u\equiv 1}.
\label{CorrFn}
\end{eqnarray}
Thus the $m^\mathrm{th}$ correlation function is obtained as a general $m^\mathrm{th}$ functional derivative of the generating functional.

Consider using this to find the probability density $X_1(p;t) \equiv X(p;t)$ for the death-diffusion model of Eq. (\ref{DeathLiouv}). From the actions in Eq. (\ref{ActsFock}) and the evolution operator expansion in Eq. (\ref{DeathEvo}) we find generating functional,
\begin{eqnarray}
<\xi(u),U\xi(v)>_\Gamma & = & \prod_{\ddd p \in \Upsilon}\left( 1+u(e^{-tH}-1)v\hh\dd p - v(e^{-tH}1-1)\hh\dd p \right) <\xi(u),\xi(v)>_\Gamma
\nonumber \\
& = & \exp\left\{ \int_\Upsilon \left(ue^{-tH}v - v(e^{-tH}1-1)\right) \hh\dd p\right\}.
\nonumber
\end{eqnarray}
If we further assume that $\mu(p) = \mu$ is also constant, note that $e^{-tH}1 = e^{-\mu t}$. Furthermore, if $V = e^{t\Delta}v$ we find $\frac{\partial V}{\partial t} = \Delta V$ with $V(p;0)=v(p)$. That is, it satisfies the heat equation and so $V(p,t) = \int_\Upsilon \dd p\hh \Phi(p-q;t)v(q)$ where $\Phi(p;t) = \frac{1}{(4\pi D t)^{d/2}}\exp(\frac{-|p|^2}{4D t})$ is the standard heat equation solution for a point source, and $d=\dim(\Upsilon)$. The generating functional then becomes,
\begin{equation}
<\xi(u),U\xi(v)>_\Gamma e^{-\int_\Upsilon v(p) \hh\ddd p} = \exp\left\{ e^{-\mu t}\iint_{\Upsilon^2} u(p) \Phi(p-q;t) v(q) \hh \dd p \hh \dd q - e^{-\mu t}\int_\Upsilon v \hh \dd p \right\}.
\label{DeathGenFn}
\end{equation}
Note this can also be derived via path integral techniques (see Appendix A).

Then from Eq. (\ref{CorrFn}), the density $X(p;t)$ has weighted integral,
\begin{equation}
\int_\Upsilon f(p) X(p;t) \hh \dd p = \left.
\frac{\partial}{\partial \epsilon}<\xi(u+\epsilon f),U\xi(v)>_\Gamma\right|_{\substack{u\equiv 1 \\ \epsilon = 0}}e^{-\int_\Upsilon v(p) \hh \ddd p} 
= e^{-\mu t}\iint_{\Upsilon^2} f(p)\Phi(p-q;t)v(q) \hh \dd p \hh \dd q,
\nonumber
\end{equation}
resulting in density,
\begin{equation}
X(p;t) = e^{-\mu t}\int_\Upsilon \Phi(p-q;t)v(q) \hh \dd p.
\nonumber
\end{equation}
Thus we find initial density $v(p)$ undergoing diffusion, down-weighted by death factor $e^{-\mu t}$, as expected. 
%%%%%%%%%%%%%%%%%%%%%%%%%%%%%%%%%%%%%%%%%%%%%%%%%%%%%%%%%%%%%

\subsection{Probability Density}

The probability density in Eq. (\ref{DensEx}) can also be interpreted in a similar manner. In the Doi formalism we find that the probability density is given by,
\begin{equation}
f^{(n)}({\bf p}_n;t) = \braket{{\bf p}_n|U|v}e^{-\int_\Upsilon v \hh \ddd p}.
\nonumber
\end{equation}
Thus the bra $\bra{{\bf p}_n} = \bra{\phi}a_{p_1}\dots a_{p_n}$ needs interpretation. To this end, we utilise the correspondence,
\begin{equation}
A^\dag_{f_1}\dots A^\dag_{f_n} \equiv \iint_{\Upsilon^n}f_1(p_1)\dots f_n(p_n)a^\dag_{p_1}\dots a^\dag_{p_n} \hh \dd {\bf p}_n.
\nonumber
\end{equation}
Then the probability density can be found via this formalism in much that same way as Eq. (\ref{CorrFn}),
\begin{eqnarray}
& & \iint_{\Upsilon^n}f_1(p_1)\dots f_n(p_n) f^{(n)}({\bf p}_n;t) \hh \dd {\bf p}_n = <A_{f_1}^\dag\dots A_{f_m}^\dag \xi(0),U\xi(v)>_\Gamma e^{-\int_\Upsilon v(p) \hh \ddd p}
\nonumber \\
& = & \lim_{{\bm{\epsilon}} \rightarrow 0} \frac{\partial^n}{\partial \epsilon_1\dots \partial\epsilon_n}<\xi({\bm{\epsilon}}. {\bf f}),U\xi(v)>_\Gamma e^{-\int_\Upsilon v \hh \ddd p}
\nonumber \\
& = & \left.\iint \dd{\bf p}_n \hh f_1(p_1)\dots f_n(p_n)\frac{\delta^n}{\delta u(p_1)\dots \delta u(p_n)} <\xi(u),U\xi(v)>_\Gamma e^{-\int_\Upsilon v(p) \hh \ddd p}\right|_{u=0}.
\nonumber
\end{eqnarray}
Thus the weighted probability density can also be written as a functional derivative. If we apply this to the death-diffusion model above, for example, substituting the generating functional from Eq. (\ref{DeathGenFn}) we get,
\begin{equation}
f^{(n)}({\bf p}_n;t) = \exp\left\{ -e^{-\mu t}\int_\Upsilon v(p) \hh \dd p\right\}
\prod_{i=1}^n \left(\int_\Upsilon v(p)\Phi(p-p_i;t)\hh \dd p\right).
\nonumber
\end{equation}

%%%%%%%%%%%%%%%%%%%%%%%%%%%%%%%%%%%%%%%%%%%%%%%%%%%%%%%%%%%%%
%%%%%%%%%%%%%%%%%%%%%%%%%%%%%%%%%%%%%%%%%%%%%%%%%%%%%%%%%%%%%

\section{Applications to Local Processes}

Now consider processes where interactions are local. That is, where effects occur at a single position. This could be creation of a daughter particle at the same location as a parent particle, or direct collision between two annihilating particles. For many such systems, the generating function can be calculated exactly using the expansion methods of the previous section, as are now shown with several examples. These include a Brownian tree process (to show non-associativity of It\^{o} multiplication),  the conversion reaction $A+A \rightarrow B$ (for multi-species interactions), the reversible reaction $A \leftrightarrow \phi$ (to examine a model with time dependent rates), and a classical death processes (to look at a discrete, non-spatial model).

%%%%%%%%%%%%%%%%%%%%%%%%%%%%%%%%%%%%%%%%%%%%%%%%%%%%%%%%%%%%%

\subsection{Brownian Tree Process}

Consider a set of particles undergoing diffusion, such that fission into daughter particles can also occur. That is, we have the process $A\rightarrow A+A$. Then the Doi operator takes the form of Eq. (\ref{BirthDiffLiouv}), which has the following formulation using Fock space operators,
\begin{equation}
\mathcal{L} = \int_\Upsilon \mu(p)\left((a_p^\dag)^2a_p - a_p^\dag a_p\right) + D(p)a_p^\dag\Delta a_p \hh \dd p
\equiv \int_\Upsilon (\dd B^{(2)}_\mu - \dd \Lambda_H)
\nonumber
\end{equation}
where $H = \mu-D\Delta$. The requisite It\^{o} multiplication can be found from  Eq. (\ref{BProdForm}). First then, the evolution operator becomes, using the independence of differentials at different positions,
\begin{equation}
U = e^{t\mathcal{L}} = \prod_{\ddd p \in \Upsilon} \left(1 + \sum_{n=1}^\infty \frac{t^n}{n!}(\dd B^{(2)}_\mu - \dd \Lambda_H)^n\right).
\label{BirthDiffEvo}
\end{equation}

Next we use the non-associative multiplication rule of Eq. (\ref{BProdForm}), in which multiplication occurs from the right. Note in particular that,
\begin{equation}
\dd B^{(2)}_{G_1}\hh \dd B^{(2)}_{G_2}\hh \dots \dd B^{(2)}_{G_n}\hh
= n!\dd B^{(n+1)}_{G_1G_2\dots G_n}.
\nonumber
\end{equation}
Also, noting that $\Lambda_H \equiv B_H^{(1)}$, and from $\dd \Lambda_G \hh \dd B^{(k)}_H  = k\dd B^{(k)}_{GH}$, inserting any $\dd \Lambda_H$ operators into the above factor results in a factor $k$ appearing at the position it was inserted (counting from the right). Thus, for example, $\dd B^{(2)}_\mu \hh \dd B^{(2)}_\mu \hh\ \dd\Lambda_H \hh \dd B^{(2)}_\mu \hh \dd B^{(2)}_\mu = 3.4!\dd B^{(5)}_{\mu^2H\mu^2}$. To further simplify things, we next assume that $\mu$ and $H$ commute. This can be achieved by assuming $\mu(p) = \mu$ is constant (or taking $\mu$ as a harmonic function, with $\Delta\mu = 0$).

Then when expanding Eq. (\ref{BirthDiffEvo}), the $\dd\Lambda_H$ terms pick up factors corresponding to their positions amongst the $\dd B_\mu^{(2)}$ operators. If we have $n-k$ operators $\dd \Lambda_H$ and $k$ operators $\dd B^{(2)}_\mu$ this results in a factor of the form (see \cite{Butler2010}),
\begin{equation}
\sum_{1 \le m_1 \le m_2 \le \dots \le m_{n-k} \le k+1} m_1m_2\dots m_{n-k} = {n+1 \brace k+1},
\nonumber
\end{equation} 
where the latter term is a Stirling number of the second kind, which can be defined via exponential generating function \cite{Stanley2011},
\begin{equation}
\sum_{n \ge k}{n \brace k}\frac{x^n}{n!} = \frac{1}{k!}(e^x-1)^k.
\label{SNSCEGF}
\end{equation}

Then, the evolution operator reduces to,
\begin{equation}
U = \prod_{\ddd p \in \Upsilon}\left(1+
\sum_{n=1}^\infty\frac{t^n}{n!}\sum_{k=0}^nk!(-1)^{n-k}{n+1 \brace k+1}
\dd B^{(k+1)}_{\mu^kH^{n-k}}
\right).
\nonumber
\end{equation}

Now pairs of terms in the product corresponding to distinct positions commute. Then noting that $<\xi(u),\dd B^{(r)}_{\mu^\alpha H^\beta}\xi(v)> = u^r\mu^\alpha H^\beta v \hh\dd p<\xi(u),\xi(v)>$, we obtain generating functional,
\begin{eqnarray}
&& <\xi(u),U\xi(v)> 
\nonumber \\
& = & <\xi(u),\prod_{\ddd p \in \Upsilon}\left(
1+\dd p\hh\sum_{n=1}^\infty \frac{t^n}{n!}\sum_{k=0}^{n}k!(-1)^{n-k}{n+1 \brace k+1}
u^{k+1}\mu^kH^{n-k}v
\right)\xi(v)>
\nonumber \\
& = & \exp \int_\Upsilon \dd p \hh \left\{
uv + \sum_{n=1}^\infty \frac{t^n}{n!}\sum_{k=0}^nk!(-1)^{n-k}{n+1 \brace k+1}
u^{k+1}\mu^kH^{n-k}v
\right\}
\nonumber \\
& = & \exp \int_\Upsilon \dd p \hh \left\{
\sum_{k=0}^\infty u^{k+1}k!(-\mu)^k
\sum_{n=k}^\infty \frac{(-t)^n}{n!}H^{n-k}{n+1 \brace k+1}v
\right\}.
\nonumber
\end{eqnarray}
Finally, using the differential of Eq. (\ref{SNSCEGF}), along with shorthand form $e_1(H,t) = H^{-1}(e^{-tH}-1)$, this reduces to,
\begin{equation}
<\xi(u),U\xi(v)> = \exp \int_\Upsilon \dd p \hh \left\{ u\sum_{k=0}^\infty u^ke^{-tH}(-\mu e_1(H,t))^kv \right\}
= \exp \int \dd p \hh \left\{ \frac{ue^{-tH}}{1+\mu ue_1(H,t)}v \right\},
\nonumber
\end{equation}
where the last term is written in the formal sense, with $H$ only acting on $v$. Features of interest can then be extracted. For example, using Eq. (\ref{CorrFn}) the density becomes,
\begin{equation}
X(p;t) = \sum_{k=0}^\infty (k+1) e^{-tH}(-\mu e_1(H,t))^kv,
\nonumber
\end{equation}
which for the static case ($D = 0$), just reduces to $X(p;t) = v(p)e^{\mu t}$ as would be expected for a distribution $v(p)$ undergoing growth at rate $\mu$.

%%%%%%%%%%%%%%%%%%%%%%%%%%%%%%%%%%%%%%%%%%%%%%%%%%%%%%%%%%%%%

\subsection{Multispecies}

Multispecies can be analysed using these techniques. Field theory approaches with two species have found applications in binary fission processes, for example \cite{Greenman2017}, as well as reactions such as $A+A \rightarrow B$ \cite{Cardy2006, Mcquarrie1964}. This involves two classes of commuting creation and annihilation operators, which can combine in a single Liouvillian. 

For example, take the case of conversion from one species to another $A \rightarrow B$. For a simple demonstration we assume a static non-diffusive model. Then the Liouvillian is given by,
\begin{equation}
\mathcal{L}  = \int_\Upsilon \dd p \hh \mu(p)(b_p^\dag a_p - a_p^\dag a_p) = \int_\Upsilon (\dd M_\mu - \dd\Lambda_\mu).
\nonumber
\end{equation}

\begin{table}[h]
\begin{center}
\begin{tabular}{||c||cc||} 
\hline
& $\dd M_G$ & $\dd \Lambda_G$ \\ [0.5ex] 
\hline\hline
$\dd M_F$ & $0$ & $\dd M_{FG}$ \\ 
\hline
$\dd \Lambda_F$ & $0$ & $\dd \Lambda_{FG}$\\
\hline
\end{tabular}
\caption{Quantum stochastic process It\^{o} table for $A\rightarrow B$.}
\label{ItoTableAnn}
\end{center}
\end{table}

This results in the It\^{o} multiplication rules given in Table \ref{ItoTableAnn}. The evolution operator then becomes,
\begin{equation}
U = e^{t\mathcal{L}} = \prod_{\ddd p \in \Upsilon}\left(1 + \sum_{n=1}^\infty\frac{t^n}{n!}(\dd M_\mu - \dd\Lambda_\mu)^n  \right) = \prod_{\ddd p \in \Upsilon}\left(1+ \sum_{n=1}^\infty\frac{(-t)^n}{n!}(\dd \Lambda_{\mu^n} - \dd M_{\mu^n})\right).
\nonumber
\end{equation}

Coherent states such as $\ket{\bf u} = \ket{u_a,u_b}$ now involve functions in $\Upsilon^2$, where the $a_p$ and $b_p$ creation operators act on the first and second functions respectively. Then the generating functional becomes,
\begin{equation}
<\xi({\bf u}), U\xi({\bf v})>_\Gamma = \exp \int_\Upsilon \dd p \hh \left\{ u_av_a+u_bv_b + (e^{-\mu t}-1)(u_a-u_b)v_a \right\}.
\nonumber
\end{equation}

Features of interest, such as the densities of the two species, can then be extracted. From Eq. (\ref{CorrFn}) we find (with a similar expression for $X_b(p;t)$),
\begin{equation}
X_a(p;t) = \left.\frac{\delta}{\delta u_a(p)}<\xi({\bf u}),U\xi({\bf v})>_\Gamma e^{-\int_\Upsilon v_a(p)+v_b(p) \hh \ddd p}\right|_{u\equiv 1}.
\nonumber
\end{equation}
This gives $X_a(p,t) = v_a(p)e^{-\mu t}$ and $X_b(p,t) = v_b(p)+v_a(p)(1-e^{-\mu t})$, describing the expected decay and transfer of densities.

%%%%%%%%%%%%%%%%%%%%%%%%%%%%%%%%%%%%%%%%%%%%%%%%%%%%%%%%%%%%%

\subsection{Time Dependent Evolution}

When time dependent Liouvillians are involved, path integrals have to respect time ordering \cite{Fetter2012, Kleinert2009, Lancaster2014, Peskin2018, Tauber2014}, where we have evolution operator,
\begin{equation}
U = \overleftarrow{\mathcal{T}}\left\{e^{\int_0^t\mathcal{L}(s) \hh\ddd s} \right\}
=\overleftarrow{\mathcal{T}}\prod_{i=1}^N e^{\mathcal{L}(s_i) \hh\ddd s},
\label{Schwinger}
\end{equation}
where $s_i = i \hh \dd s$ with $\dd s = t/N$. Note that the latter product indicates how the integral in the exponent should be interpreted. We will see that this time integral can be done effectively when using quantum stochastic process techniques. The case of a single time dependent function in the Liouvillian is fairly straightforward. The case with multiple time functions is a bit more involved, but tractable. Both cases are considered below.

%%%%%%%%%%%%%%%%%%%%%%%%%%%%%%%%%%%%%%%%%%%%%%%%%%%%%%%%%%%%%

\subsubsection{A Single Time Function}

Firstly, we treat the case of one time dependent function, where time ordering will be seen to play no role. To see this, consider for example the simple model $\phi \rightarrow A$ of spontaneous creation of particles with Liouvillian,
\begin{equation}
\mathcal{L}(t) = \int_\Upsilon\mu(p,t)(a_p^\dag-1) \hh \dd p.
\nonumber
\end{equation}
Here $\mu(p,t)$ is the rate of particle production, varying in both space and time.

Note that $[\mathcal{L}(s_i),\mathcal{L}(s_j)]=0$, meaning the exponents in the time ordered product in Eq. (\ref{Schwinger}) can be added together in a single exponent, and so the integral $\int_0^t\mathcal{L}(s) \dd s$ can be calculated directly. Then if $M(p,t) = \int_0^t \mu(p,s) \hh \dd s$ denotes the cumulative rate of production, the generating functional becomes,
\begin{eqnarray}
\braket{u|U|v}e^{-\int_\Upsilon v(p)\hh\ddd p} & = & \braket{u|\overleftarrow{\mathcal{T}}
\exp\left\{ \int_0^t \dd s \int_\Upsilon \dd p \hh\mu(p;s)(a_p^\dag-1) \right\}|v} e^{-\int_\Upsilon v(p)\hh\ddd p}
\nonumber \\ 
& = & \braket{u|\exp\left\{\int_\Upsilon \dd p \hh M(p;t)(a_p^\dag-1) \right\}|v} e^{-\int_\Upsilon v(p)\hh\ddd p}
\nonumber \\
& \equiv & <\xi(u),e^{\int_\Upsilon \ddd A^\dag_M}\xi(v)>_\Gamma e^{-\int_\Upsilon (v(p)+M(p,t))\hh\ddd p}
\nonumber \\
& = & <\xi(u),\prod_{\ddd p \in \Upsilon}(1+\sum_{n=1}^\infty \frac{(\dd A^\dag_M)^n}{n!})\xi(v)>_\Gamma e^{-\int_\Upsilon (v(p)+M(p,t))\hh\ddd p}
\nonumber \\
& = & <\xi(u),\prod_{\ddd p \in \Upsilon}(1+\dd A^\dag_M)\xi(v)>_\Gamma e^{-\int_\Upsilon (v(p)+M(p,t))\hh\ddd p}
\nonumber \\
& = & \exp\left\{\int_\Upsilon (v(p)+M(p,t))(u(p)-1)\hh \dd p\right\}.
\nonumber
\end{eqnarray}
Note that the only It\^{o} product needed in the expansion is $(\dd A_M^\dag)^2=0$. Then from Eq. (\ref{CorrFn}), the particle density becomes $X(p;t)=M(p,t)+v(p)$, with the cumulative time dependent rate of spontaneous birth adding to the initial distribution, as expected. 

%%%%%%%%%%%%%%%%%%%%%%%%%%%%%%%%%%%%%%%%%%%%%%%%%%%%%%%%%%%%%

\subsubsection{A Pair of Time Functions}

Secondly then, we treat a case where two time dependent functions are present. The previous model is extended, and we now consider the combined dual processes $\phi \rightarrow A$ and $A \rightarrow \phi$ of spontaneous birth and death, with Liouvillian,
\begin{equation}
\mathcal{L}(t) = \int_\Upsilon [\mu(p,t)(a_p^\dag-1) + \nu(p,t)(a_p-a_p^\dag a_p)]\hh\dd p.
\nonumber
\end{equation}

We now find that $[\mathcal{L}(s_i),\mathcal{L}(s_j)] \ne 0$ (for distinct times $s_i$ and $s_j$), and we cannot simply add the exponents to deal with the integral form of Eq. (\ref{Schwinger}), but have to treat the time ordered product directly. Although this product form is used to construct path integrals (which would be one valid approach), it has also been used in Schwinger's prescription for time dependent systems, where time ordered products of the form $e^{X_N} e^{X_{n-1}} \dots e^{X_2} e^{X_1}$ are converted into $e^{X_N \diamond \dots \diamond X_2 \diamond X_1}$, where $\diamond$ indicates the utilisation of Baker-Campbell-Hausdorff techniques to combine the terms into a single exponent \cite{Mielnik1970}. We now adopt a similar approach, except now the quantum stochastic process machinery is applied to combine the terms.

To start, the evolution operator can be written as follows, where we use shorthand $\nu_i = \nu(\cdot,s_i)$ and $\mu_i = \mu(\cdot,s_i)$ for functions acting on space $\Upsilon$ for a given value $s_i$ of time,
\begin{eqnarray}
U(t) & = &  \overleftarrow{\mathcal{T}}\prod_{i=1}^N e^{\ddd s \hh \mathcal{L}(s_i)}
=  \overleftarrow{\mathcal{T}}\prod_{i=1}^N e^{\ddd s \hh (X_{\nu_i}+Y_{\mu_i})}
= \overleftarrow{\mathcal{T}}\prod_{i=1}^N 
\exp \left\{ \dd s\int_\Upsilon (\dd X_{\nu_i} + \dd Y_{\mu_i}) \right \} 
\nonumber \\
& = & \overleftarrow{\mathcal{T}}\prod_{i=1}^N 
\exp \left\{ \dd s\sum_{\ddd p \in \Upsilon} (\dd X_{\nu_i} + \dd Y_{\mu_i}) \right \} 
= \overleftarrow{\mathcal{T}}\prod_{i=1}^N \prod_{\ddd p \in \Upsilon} 
\left(1+ \dd s(\dd X_{\nu_i} + \dd Y_{\mu_i})\right).
\label{TimeDepProd}
\end{eqnarray}
Note that order $\dd s^2$ terms (or higher) in the product do not survive, and we have operators,
\begin{eqnarray}
X_{\nu_i} & = & \int_\Upsilon \mu(p,s_i) (a_p^\dag-I) \hh \dd p
\equiv \int_\Upsilon (\dd A^\dag_{\mu_i} - \mu(p,s_i)\hh\dd p) = \int_\Upsilon \dd X_{\nu_i}.
\nonumber\\
Y_{\mu_i} & = & \int_\Upsilon \nu(p,s_i) (a_p-a_p^\dag a_p) \hh \dd p 
\equiv \int_\Upsilon (\dd A_{\nu_i} - \dd \Lambda_{\nu_i}) = \int_\Upsilon \dd Y_{\mu_i}.
\nonumber
\end{eqnarray}

Multiplication rules for $\dd X_\nu$ and $\dd Y_\mu$ can be found in Table \ref{ItoTableDual}.
\begin{table}[t!]
\begin{center}
\begin{tabular}{||c||cc||} 
\hline
& $\dd X_{\nu'}$ & $\dd Y_{\mu'}$ \\ [0.5ex] 
\hline\hline
$\dd X_\nu$ & $0$ & $0$ \\ 
\hline
$\dd Y_\mu$ & -$\dd X_{\mu\nu'}$ & -$\dd Y_{\mu\nu'}$\\
\hline
\end{tabular}
\caption{Quantum stochastic process It\^{o} table for joint process $\phi \rightarrow A$ and $A \rightarrow \phi$.}
\label{ItoTableDual}
\end{center}
\end{table}
Now, the terms $\dd X_{\nu_i} \equiv \dd X_{\nu(p,s_i)}$ and $\dd Y_{\mu_i}\equiv \dd Y_{\mu(p,s_i)}$ commute for distinct $p$, so terms from the product in Eq. (\ref{TimeDepProd}) corresponding to a particular interval $[p,p+\dd p]$ can be collected together, in time ordered sequence. Then, we have $U = \prod_{\ddd p \in \Upsilon}U_p$, where $U_p$ is given as follows (with the aid of Table \ref{ItoTableDual}),
\begin{eqnarray}
U_p & = & \overleftarrow{\mathcal{T}}\prod_{i=1}^N 
\left(1 + \dd s(\dd X_{\nu_i} + \dd Y_{\mu_i})\right),
\nonumber \\
& = & 1 + \dd s \sum_{N \ge i \ge 0}(\dd X_{\mu_i} + \dd Y_{\nu_i})
-\dd s^2 \sum_{N \ge i > j \ge 0}(\dd X_{\nu_i\mu_j}+ \dd Y_{\nu_i\nu_j})
\nonumber \\
&& +\dd s^3 \sum_{N \ge i > j > k \ge 0}(\dd X_{\nu_i\nu_j\mu_k}+ \dd Y_{\nu_i\nu_j\nu_k})
- \dots.
\nonumber
\end{eqnarray}

Then, noting the actions from Eq.s (\ref{ActsFock}) (for general function $\alpha$),
\begin{eqnarray}
<\xi(u),\dd X_\alpha(q)\xi(v)>_\Gamma & = & <\xi(u),\xi(v)>_\Gamma\dd q \hh \int_\Upsilon \alpha(q,t)(u_q-1),
\nonumber \\
<\xi(u),\dd Y_\alpha(q)\xi(v)>_\Gamma  & = & <\xi(u),\xi(v)>_\Gamma\dd q \hh \int_\Upsilon \alpha(q,t)(1-u_q)v_q,
\nonumber
\end{eqnarray}
leads to the following inner product (using shorthand such as $\mu(p,s)\equiv\mu_p(s)$ and $u_p \equiv u(p)$),
\begin{eqnarray}
\frac{<\xi(u),U_p\xi(v)>_\Gamma}{<\xi(u),\xi(v)>_\Gamma} & = & 1 + \dd p(u_p-1) \hh \left[\int_0^t \mu_p(s) \dd s
-\int_0^t \mu_p(s)\dd s \hh \int_s^t \nu_p(\tau) \dd \tau\right.
\nonumber \\
&& \hspace{3cm} + \left.\int_0^t \mu_p(s)\dd s \hh \int_{t \ge \tau_2 \ge \tau_1 \ge s} \nu_p(\tau_2)\nu_p(\tau_1) \dd \tau
+ \dots\right]
\nonumber \\
&& + \dd p \hh(1-u_p)v_p\left[ \int_0^t \nu_p(s) \dd s
-\int_{t \ge \tau_2 \ge \tau_1 \ge 0} \nu_p(\tau_2)\nu_p(\tau_1) \dd \tau + \dots
\right]
\nonumber \\
& = & 1 + \dd p \hh (u_p-1)\int_0^t\mu_p(s) e^{-\int_s^t \nu(\tau) \ddd \tau}\dd s \hh
+\dd p \hh(1-u_p)v_p\left(1-e^{-\int_0^t\nu_p(s)\ddd s}\right).
\nonumber
\end{eqnarray}

Finally, noting the actions of $U_p$ commute for distinct $p$, we finally get the generating functional,
\begin{equation}
G(u,v) = \braket{u|Uv}e^{-\int_\Upsilon v_p \ddd p} = 
\exp\left\{ \int_\Upsilon \dd p \hh 
\left[(u_p-1)\int_0^t\mu_p(s) e^{-\int_s^t \nu(\tau) \ddd \tau}\dd s \hh
+ (u_p-1)v_q e^{-\int_0^t\nu_p(s)\ddd s}\right]
\right\}.
\nonumber
\end{equation}

The particle density is then found to be,
\begin{equation}
X(p;t) = v(p)e^{-\int_0^t \nu_p(s)\ddd s} + \int_0^t \mu_p(s)e^{-\int_s^t \nu_p(\tau) \ddd \tau},
\nonumber
\end{equation}
representing death of the initial distribution, and birth-death interaction, as expected. 

%%%%%%%%%%%%%%%%%%%%%%%%%%%%%%%%%%%%%%%%%%%%%%%%%%%%%%%%%%%%%

\subsection{Discrete Models}

Lastly, discrete models are considered. These are models such as classic birth and death, where only the population size is of concern, there are no spatial dependencies. Results can be obtained by making the spatially dependent functions of previous sections uniform, and removing spatial kinetics such as diffusion. For example, a simple death model can be written with Liouvillian $\mathcal{L} = \mu(a-a^\dag a)$ where $a$ and $a^\dag$ are standard annihilation and creation operators with commutator $[a,a^\dag] = 1$. This is the non-spatial version of Eq. (\ref{DeathLiouv}). Then if we assume a population initially Poisson distributed with mean $v$, the generating function is given by,
\begin{equation}
\braket{u|e^{t\mathcal{L}}|v}e^{-v} \equiv <\xi(u),e^{\mu t (\ddd A - \ddd \Lambda)}\xi(v)>_\Gamma e^{-v}
 = <\xi(u),(1+\sum_{n=1}^\infty\frac{(\mu t)^n}{n!}(\dd A - \dd \Lambda)^n) \xi(v)>_\Gamma e^{-v}.
\nonumber
\end{equation} 
Note that the coherent states are no longer of the continuous variety found in Doi \cite{Doi1976a, Doi1976b}, but are of the discrete form found in most applications, such as those used by Peliti \cite{Peliti1985}, for example.

Then from the quantum stochastic It\^{o} products of Table \ref{ItoTable} we get the spatially uniform relationships $\dd A \hh \dd \Lambda = \dd A$ and $\dd \Lambda^2 = \dd \Lambda$, with all other relevant products being zero. Then using these relationships along with the spatially uniform operations $<\xi(u),\dd A\hh\xi(v)>_\Gamma = v<\xi(u),\xi(v)>_\Gamma$ and $<\xi(u),\dd \Lambda\hh \xi(v)>_\Gamma = uv<\xi(u),\xi(v)>_\Gamma$ and we end up with,
\begin{equation}
\braket{u|U|v}e^{-v} = e^{(u-1)ve^{-\mu t}}.
\nonumber
\end{equation}

This is the spatially uniform version of Eq. (\ref{DeathGenFn}). However, this was simply derived above by ignoring spatial aspects and extracting the algebraic properties from the It\^{o} multiplication table, giving a simpler realization than the Evans-Hudson flows in \cite{Fagnola1991, Parthasarathy1990}. Note that this generating function can also be simply derived by classical generating function methods \cite{Jones2017}.

Features of interest are then readily derivable. The exponential decay of mean population size is, for example, $\left.\frac{\ddd}{\ddd u}\braket{u|U|v}e^{-v}\right|_{u=1} = ve^{-\mu t}$, as expected.

%%%%%%%%%%%%%%%%%%%%%%%%%%%%%%%%%%%%%%%%%%%%%%%%%%%%%%%%%%%%%
%%%%%%%%%%%%%%%%%%%%%%%%%%%%%%%%%%%%%%%%%%%%%%%%%%%%%%%%%%%%%

\section{Non-Local Processes and Perturbation}

Processes involving non-local interactions are the most general found in the Doi framework \cite{Doi1976a, Doi1976b}, and although they can be represented in the It\^{o} framework as described below, they do not replace perturbation expansion methods, which are still required. 

Consider for example the annihilation process $A+A \rightarrow \phi$, which has been studied (along with the more general case $kA \rightarrow \phi$) as a benchmark model for field theory techniques in reaction diffusion systems \cite{Cardy2006, Cardy1998, Doi1976a, Doi1976b, Hnatich2000, Lee1994, Tauber2014}. They can also be viewed as a simplification of the conversion process $A+A \rightarrow B$ where the density of $B$ particles is not of interest. 

There are two natural approaches to this. One is the continuous Doi approach, typically used when interactions operate over a distance. The original work of Doi, for example \cite{Doi1976b}, treated this model with annihilation occurring at rate $R_r$ for particles separated by distance $r$. In many applications interactions are local and a lattice model is used \cite{Peliti1985, Tauber2014}, whereby annihilations occur on lattice sites and scaling methods are used to go to the continuum. 

We next consider how these approaches can be treated with stochastic quantum process techniques. Specifically, we firstly see how the product expansion methods of previous sections do not directly work for non local interactions, and secondly, see how perturbations can instead  be formulated. 

%%%%%%%%%%%%%%%%%%%%%%%%%%%%%%%%%%%%%%%%%%%%%%%%%%%%%%%%%%%%%

\subsection{Doi Liouvillian, Lattice Methods and Diffusion Limited Reactions}

In continuous Doi formulation, the Liouvillian operator for process $A+A\rightarrow\phi$ takes the following form, where we assume that the diffusion rate $D$ is fixed.
\begin{equation}
\mathcal{L} = \frac{1}{2}\iint_{\Upsilon^2} R_{p-q}\left(a_pa_q - a_p^\dag a_q^\dag a_pa_q\right) \hh \dd p \hh \dd q
+ \int_\Upsilon Da_p^\dag \Delta a_p \hh \dd p 
\equiv \iint_{\Upsilon^2}(\dd\hh\Omega_R - \dd\Xi_R) + \int_\Upsilon\dd\Lambda_{D\Delta}.
\label{AnnihDiffLiouv}
\end{equation}
The factor $\frac{1}{2}$ reflects the fact that swapping $p$ and $q$ represents the same event.

It is worth pointing out that the model $A+A\rightarrow\phi$ has frequently been modelled using discrete lattice methods (see Section \ref{Lattice}). These primarily have a model where particle pairs annihilate when they meet on the same lattice site. If we use the reaction rate $R_{p-q} = R\delta(p-q)$, the Liouvillian in Eq. (\ref{AnnihDiffLiouv}) becomes,
\begin{equation}
\mathcal{L} = \frac{R}{2}\int_\Upsilon \left(a_p^2 - (a_p^\dag)^2a_p^2\right) \hh \dd p
+ \int_\Upsilon Da_p^\dag \Delta a_p \hh \dd p.
\nonumber
\end{equation}
This is precisely the form seen using lattice methods after the lattice constant $\epsilon \rightarrow 0$ goes to the continuum limit. It would seem that in the limit particles are now unlikely to ever meet, even if they annihilate with certainty whenever they do. However, note that if two particles have spatial distributions $v_p$ then the annihilation rate is given by $\frac{1}{2}\iint_{\Upsilon^2} \dd p \hh \dd q \hh R_{p-q}v_pv_q = \frac{R}{2}\int_\Upsilon \dd p \hh v_p^2$, and we find a positive annihilation rate. Thus we have a model where annihilation is immediate upon contact, and where particles collide with positive probability. This is often referred to as a diffusion limited system, used to model cases where the reaction rate is much higher than the diffusion rate. Note that the formulation in Eq. (\ref{AnnihDiffLiouv}) is more general than this, and allows for reactions that are not diffusion limited. 
 
%%%%%%%%%%%%%%%%%%%%%%%%%%%%%%%%%%%%%%%%%%%%%%%%%%%%%%%%%%%%%

\subsection{Limitations of Previous Methods}

Some difficulties arising from the methods used in previous sections now become apparent. Firstly, the operators in Eq. (\ref{AnnihDiffLiouv}) are acting in distinct spaces $\Upsilon^2$ and $\Upsilon$. It is tempting to replace the third term with expression $\iint_{\Upsilon^2} Dh(q)a_p^\dag \Delta a_p \hh \dd p \hh \dd q$, utilising any dummy function $h(q)$ with unit integral. The multiplication table for the three operators could then be constructed. However, this would not solve the problem. Assume $D=0$ for a moment. Now, attempting to write the evolution operator as a product reveals a second problem,
\begin{equation}
U = e^{t\mathcal{L}} = \prod_{(\ddd p,\ddd q) \in \Upsilon^2}\left(1+\sum_{n=1}^\infty\frac{t^n}{n!}(\dd \Omega_R - \dd \Xi_R)^n\right).
\label{WrongProd}
\end{equation}
In previous examples, where the Liouvillian $\mathcal{L} = \int_\Upsilon \dd S(p)$ is over space $\Upsilon$, we find that $\dd S(p)$ and $\dd S(q)$ commute for distinct $p$ and $q$ (because the commutator relation $[a_p,a_q^\dag]=\delta(p-q)$ is not implicated when corresponding intervals $[p,p+\dd p]$ and $[q,q+\dd q]$ do not overlap), meaning the evolution operator $U$ can be broken up into a product over $\dd p$. However, in the case of Eq. (\ref{WrongProd}) the action is over space $\Upsilon^2$ and we find that $\dd S(p,p')$ and $\dd S(q,q')$ no longer commute for distinct pairs $(p,p')$ and $(q,q')$, because, for example, we can have $p = q$ and $p' \ne q$. The product expression in Eq. (\ref{WrongProd}) is therefore invalid. Using techniques such as Zassenhaus formulae to correctly form products will likely not be helpful (even if it were tractable), as individual terms in such a product would still not commute and finding terms such as $\braket{u|Uv}$ would prove difficult.

One is left then with more standard perturbation methods. Path integrals are one choice, which circumvents the entire It\^{o} formalism. Alternatively, there is the interaction picture, the method employed by Doi for this model \cite{Doi1976b, Fetter2012}. We take this approach and show that perturbation expansion can be phrased within the It\^{o} formalism.

%%%%%%%%%%%%%%%%%%%%%%%%%%%%%%%%%%%%%%%%%%%%%%%%%%%%%%%%%%%%%

\subsection{Perturbative Approach}

In Doi's original formulation \cite{Doi1976b}, a Dyson time series expansion method is used without using what is now known as a Doi shift (see below, or \cite{Greenman2017, Peliti1985, Tauber2014} for example). This means the quadratic and quartic terms in the Liouvillian of Eq. (\ref{AnnihDiffLiouv}) result in Feynman diagrams containing nodes of degrees two and four. The analysis in \cite{Doi1976b} was also carried out in \emph{momentum space} (that is, the Fourier transform of the original \emph{position space}). For diffusive models, using momentum space has the advantage of turning the Laplacian operator into a function which are easier to handle. In later analyses, such as \cite{Cardy2006, Tauber2014} where this model (and more general models such as $mA \rightarrow nA$) are analysed, path integrals are used. These employ a Doi shift which results in a modified Liovillian and the subsequent diagrams contain cubic and quartic nodes. These analyses utilise a mixture of position and momentum space, and tend to be focused on diffusion limited local reactions. In what follows we will use the Doi shift in momentum space, with non-local reactions, which has not been presented elsewhere. We will also frame it using quantum It\^{o} calculus techniques. 

The role quantum stochastic process operators play occurs in the earlier stages of perturbation, after which methods become more standard. These will be explained in turn. Firstly, its role in the Doi shift shall be considered. Secondly, the validity of It\^{o} calculus in momentum space is justified. Thirdly, we see calculation of propagators within the interaction picture framework. Fourthly, properties of perturbative expansion will be derived. It is the intention of this section to highlight roles the It\^{o} formalism can play, rather than provide an extensive perturbation analysis for the model $A+A\rightarrow \phi$, and only an overview of perturbation is provided.

%%%%%%%%%%%%%%%%%%%%%%%%%%%%%%%%%%%%%%%%%%%%%%%%%%%%%%%%%%%%%

\subsubsection{Doi Shift}

Now, we have a generating functional of the form,
\begin{equation}
G(u,v) = \braket{u|U|v}e^{-\int_\Upsilon v_p\hh \ddd p}
=\braket{\phi|e^{\int_\Upsilon \ddd p \hh u_pa_p}U|v}e^{-\int_\Upsilon v_p \ddd p}.
\nonumber
\end{equation}
The aim of the Doi shift is to push the factor $e^{\int_\Upsilon \ddd p \hh u_pa_p}$ to the right of the evolution operator $U$. This has a notable advantage in perturbation analysis, as the number of left arms in Feynman diagrams is substantially reduced. To this end, noting that $\int_\Upsilon \dd p \hh u_pa_p \equiv \int_\Upsilon \dd A_u$, we find,
\begin{eqnarray}
e^{\int_\Upsilon \ddd A_u}\dd A_v^\dag(q) & = & \prod_p e^{\ddd A_u}\dd A_v^\dag(q)
= \prod_p (1+\dd A_u(p))\dd A_v^\dag(q)
\nonumber\\
& = & (\dd A_v^\dag(q)+\dd q \hh u_qv_q)\prod_{p\ne q} (1+\dd A_u(p))
= (\dd A_v^\dag(q)+\dd q \hh u_qv_q)\prod_p (1+\dd A_u(p))
\nonumber\\
& = & (\dd A_v^\dag(q)+\dd p \hh u_qv_q)e^{\int_\Upsilon \ddd A_u}.
\nonumber
\end{eqnarray}
Formulated in terms of Doi operators, this tells us that if $e^{\int_\Upsilon \ddd p \hh u_pa_p}$ acts from the left we get the shift $\int_\Upsilon \dd q \hh v_q a_q^\dag \rightarrow \int_\Upsilon \dd q\hh v_q (a_q^\dag+u_q)$. Given that $e^{\int_\Upsilon \ddd p \hh u_pa_p}$ commutes with $a_q$, we obtain the result for general operator $F(a_q^\dag,a_q)$,
\begin{equation}
e^{\int \ddd p \hh u_pa_p} F(a_q^\dag,a_q) 
= F(a_q^\dag+u_q,a_q)e^{\int \ddd p \hh u_pa_p}.
\nonumber
\end{equation}

The constant function $u\equiv 1$ proves most useful for calculating moments, and is known as the Doi shift. Specifically, expectations of $F(a_q,a_q^\dag)$ can be written as,
\begin{eqnarray}
\braket{F(a_q^\dag,a_q)}_{\psi_t} & = & \braket{1|F(a_q^\dag,a_q) e^{t\mathcal{L}(a_p,a_p^\dag)}|v} e^{-\int_\Upsilon v_p \hh \ddd p}
= \braket{\phi|e^{\int_\Upsilon \ddd p \hh a_p}F(a_q^\dag,a_q) e^{t\mathcal{L}(a_p,a_p^\dag)}|v} e^{-\int_\Upsilon v_p \hh \ddd p}
\label{MeanDoiShift}\\
& = & \braket{\phi|F(a_q^\dag+1,a_q) e^{t\mathcal{L}(a_p,a_p^\dag+1)}e^{\int_\Upsilon \ddd p \hh a_p}|v} e^{-\int_\Upsilon v_p \hh \ddd p}
= \braket{\phi|F(a_q^\dag+1,a_q) e^{t\mathcal{L}(a_p,a_p^\dag+1)}|v}.
\nonumber
\end{eqnarray}
Thus we find a simple vacuum state on the left and the normalisation factor has nicely cancelled on the right. The Doi shifted Liouvillian thus becomes,
\begin{equation}
\mathcal{L} = -\iint_{\Upsilon^2} R_{p-q}\left(a_p^\dag a_pa_q + \frac{1}{2}a_p^\dag a_q^\dag a_pa_q\right) \hh \dd p \hh \dd q
+ \int_\Upsilon Da_p^\dag \Delta a_p \hh \dd p
\equiv-\Omega_R-\Xi_R+\Lambda_{D\Delta}.
\nonumber
\end{equation}

%%%%%%%%%%%%%%%%%%%%%%%%%%%%%%%%%%%%%%%%%%%%%%%%%%%%%%%%%%%%%

\subsubsection{Momentum Space}

Next then, we consider conversion into momentum space. We start with operators,
\begin{equation}
a_k = \frac{1}{(2\pi)^{d/2}}\int_\Upsilon a_p e^{i\hh p.k} \hh \dd p,
\hs
a_k^\dag = \frac{1}{(2\pi)^{d/2}}\int_\Upsilon a_p^\dag e^{-i\hh p.k} \hh \dd p.
\nonumber
\end{equation}
The letters $k,l,m,n$ shall be used for momentum space, and $p,q,r$ for position space. This will also apply to functions, where $u_k$ will mean the Fourier transform of $u_p$, for example. We use the same symbol for these two (distinct) functions, the context of the representation (position or momentum) should make the meaning clear. We assume space $\Upsilon = \mathbb{R}^d$ with dimension $d$. Then from these definitions, standard commutation relations $[a_k,a_l^\dag] = \delta(k-l)$ follow, and the Liouvillian $\mathcal{L} = \Lambda_{D\Delta}-\Omega_R-\Xi_R$ can be transformed from the relations,
\begin{eqnarray}
\Lambda_{D\Delta} & \equiv & D \int_\Upsilon \dd p \hh a_p^\dag \Delta a_p = -D \int_\Upsilon \dd k \hh k^2 a_k^\dag a_k \equiv -\int_\Upsilon \dd\hh\Lambda_{Dk^2},
\nonumber\\
\Omega_R & \equiv & \iint_{\Upsilon^2} \dd p \hh \dd q \hh R_{p-q} a_p^\dag a_pa_q = 
\iint_{\Upsilon^2} \dd k \hh \dd l \hh R_k a^\dag_{k+l}a_k a_l,
\nonumber\\
\Xi_R & \equiv & \frac{1}{2}\iint_{\Upsilon^2} \dd p \hh \dd q \hh R_{p-q}a_p^\dag a_q^\dag a_pa_q = \frac{1}{2(2\pi)^{d/2}}\iiint_{\Upsilon^3} \dd k \hh \dd l \hh \dd m \hh 
R_ka_{l+k}^\dag a_{m-k}^\dag a_l a_m.
\nonumber
\end{eqnarray}
Note also that $\ket{u_p} = e^{\int_\Upsilon \ddd p \hh u_pa_p^\dag}\ket{\phi} = e^{\int_\Upsilon \ddd k \hh u_ka_k^\dag}\ket{\phi} = \ket{u_k}$ and coherent states for a function $u_p$ are equal to coherent states for the Fourier transformed function $u_k$.

Now, for exposition we shall restrict attention to the moments, in particular the density, where, from Eq. (\ref{MeanDoiShift}) (treating the vacuum state $\ket{\phi}=\ket{0}$ as a coherent state with constant function $0$),
\begin{equation}
X(p;t) = \braket{a_p}_{\psi_t} = \braket{1|a_pU|v}e^{-\int v_p \hh \ddd p} = \braket{0|a_pe^{t\mathcal{L}(a_p,a_p^\dag+1)}|v}. 
\nonumber
\end{equation}

At this stage path integrals could be used, but we choose to use the interaction picture, which is more amenable to the tools of It\^{o} calculus. 

%%%%%%%%%%%%%%%%%%%%%%%%%%%%%%%%%%%%%%%%%%%%%%%%%%%%%%%%%%%%%

\subsubsection{The Interaction Picture}

In the interaction picture \cite{Doi1976b, Fetter2012, Maggiore2005, Peskin2018}, the tractable part of the Liouvillian (often referred to as the free or non-interacting part) is used to transform states and operators. Then, using $\Lambda_{D\Delta}=-\Lambda_{Dk^2}$ to represent the non-interactive (diffusive) part, we have bras, kets and operators in the interaction picture defined by,
\begin{eqnarray}
\bra{u(t)}_I & = & \bra{u}e^{-t\Lambda_{Dk^2}},
\nonumber\\
\ket{u(t)}_I & = & e^{t\Lambda_{Dk^2}}\ket{u},
\nonumber\\
\mathcal{O}_I(t) & = & e^{t\Lambda_{Dk^2}}\mathcal{O}e^{-t\Lambda_{Dk^2}}.
\nonumber
\end{eqnarray}
Then in the interaction picture, the (Fourier transformed) particle density becomes,  
\begin{eqnarray}
X(k;t) & = &  \bra{0(t)}_Ia_k(t)\mathcal{\overleftarrow{T}}\exp\left\{-\int_0^t\dd s \hh \left(\Omega_R(s) +\Xi_R(s)\right) \right\}\ket{v(0)}_I
\nonumber\\
& = & \bra{0}\mathcal{\overleftarrow{T}}a_k(t)\exp\left\{-\int_0^t\dd s \hh \left(\Omega_R(s) +\Xi_R(s)\right) \right\}\exp\left\{\int_\Upsilon v_pa_p \hh \dd p\right\}\ket{0}.
\nonumber
\end{eqnarray}
Here, note that $\bra{0(t)}_I = \bra{0}$ and $\ket{v(0)}_I=\ket{v}$, so the Doi shift has the effect that the bra and ket simplify trivially resulting in a simpler expression than methods without the Doi shift (compare with Eq. (34) in \cite{Doi1976b}). This expression can now be expanded and resolved into products of propagators via Wick's theorem \cite{Doi1976b, Maggiore2005, Peskin2018}, where propagators are defined by,
\begin{equation}
G(k,l;t,s) = \braket{0|\mathcal{\overleftarrow{T}}a_k(t)a_l^\dag(s)|0}.
\nonumber
\end{equation}

Now, these can be calculated with the aid of quantum It\^{o} calculus. Firstly, note that defining,
\begin{eqnarray}
&& \dd A_k(t) = e^{t\Lambda_{Dk^2}}\dd A_ke^{-t\Lambda_{Dk^2}} 
= \prod_l e^{t\ddd\Lambda_{Dl^2}} \dd A_k \prod_m e^{-t\ddd\Lambda_{Dm^2}} 
= e^{t\ddd\Lambda_{Dk^2}} \dd A_k e^{-t\ddd\Lambda_{Dk^2}}
\nonumber\\
& = & (1+\sum_{n=1}^\infty\frac{t^n}{n!}(\dd \Lambda_{Dk^2})^n)  \dd A_k
(1+\sum_{n=1}^\infty\frac{(-t)^n}{n!}(\dd \Lambda_{Dk^2})^n)
\nonumber\\
& = & (1+\dd \Lambda_{[e^{tDk^2}-1]})\dd A_k(1+\dd \Lambda_{[e^{-tDk^2}-1]})
= \dd A_k(1+\dd \Lambda_{[e^{-tDk^2}-1]}) = \dd A_{[e^{-tDk^2}]}.
\nonumber
\end{eqnarray}
In much the same way, we find,
\begin{equation}
\dd A^\dag_l(s) = e^{s\Lambda_{Dl^2}}\dd A^\dag_le^{-s\Lambda_{Dl^2}} = \dd A^\dag_{[e^{sDl^2}]}.
\nonumber
\end{equation}
Then we find $\mathcal{\overleftarrow{T}}\dd A_k(t)\dd A^\dag_l(s) = \dd k\delta_{k,l}\theta(t-s)e^{-(t-s)Dk^2}$ from which we infer the standard result $G(k,l;t,s) = \delta_{k,l}\theta(t-s)e^{-(t-s)Dk^2}$.

The analysis from this point forward no longer requires It\^{o} calculus and is similar to methods reported elsewhere \cite{Cardy2006, Cardy1998, Doi1976a, Doi1976b, Hnatich2000, Lee1994, Tauber2014}, except that now we have interaction over a distance, which results in some differences we highlight.

%%%%%%%%%%%%%%%%%%%%%%%%%%%%%%%%%%%%%%%%%%%%%%%%%%%%%%%%%%%%%

\subsection{Perturbation}

\begin{figure}[!t]
\centering
\setlength{\unitlength}{0.1\textwidth}
\begin{picture}(10,2.5)
\put(0,0){\includegraphics[width=\textwidth]{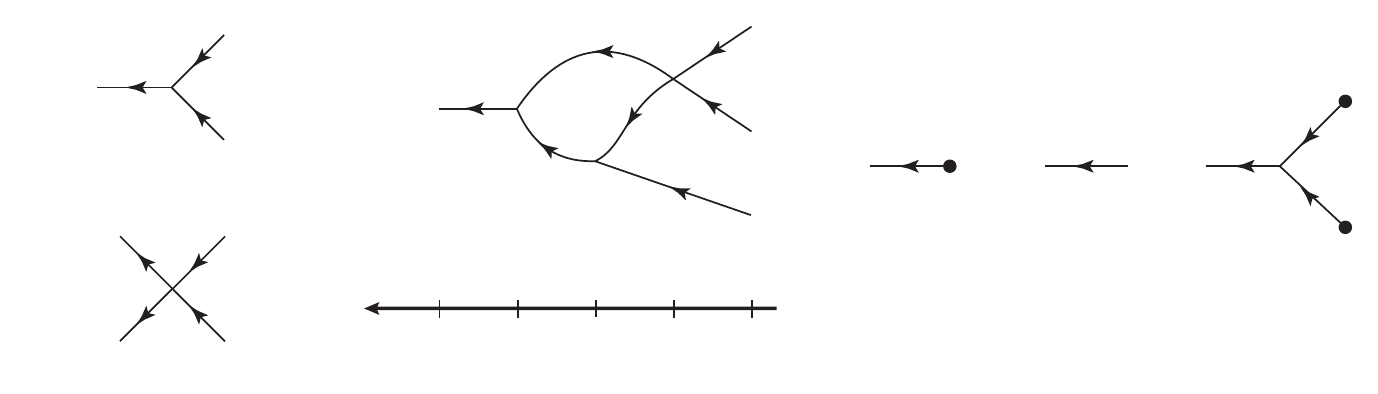}}
\put(0.20,2.60){(a)}
\put(0.85,2.30){\scriptsize $k+l$}
\put(1.55,2.35){\scriptsize $k$}
\put(1.55,2.00){\scriptsize $l$}
\put(1.00,2.00){\scriptsize $-R_k$}
\put(0.65,0.85){\scriptsize $l+k$}
\put(0.55,0.60){\scriptsize $m-k$}
\put(1.35,1.00){\scriptsize $l$}
\put(1.50,0.60){\scriptsize $m$}
\put(0.95,0.30){\scriptsize $-\frac{R_k}{2(2\pi)^d}$}
\put(2.20,2.60){(b)}
\put(2.30,0.30){Time}
\put(3.14,0.30){$t$}
\put(3.68,0.30){$\tau_3$}
\put(4.22,0.30){$\tau_2$}
\put(4.76,0.30){$\tau_1$}
\put(5.40,0.30){$0$}
\put(3.40,2.15){\scriptsize $k$}
\put(3.80,2.55){\scriptsize $k-m-n+l$}
\put(3.40,1.55){\scriptsize $m+n-l$}
\put(4.60,1.85){\scriptsize $m-l$}
\put(4.75,1.35){\scriptsize $n$}
\put(5.15,2.15){\scriptsize $m$}
\put(5.25,2.35){\scriptsize $k-m-n$}
\put(6.30,2.60){(c)}
\put(7.15,1.60){$=$}
\put(8.40,1.60){$+$}
\end{picture}
\caption{Feynman diagrams for $A+A\rightarrow \phi$ perturbative expansion. (a) Nodes corresponding to $\Omega_R$ (top) and $\Xi_R$ (bottom). (b) A third order sample diagram. (c) Recursion for tree diagrams, where each circle represents all possible tree diagrams.}
\label{PertFig}
\end{figure}
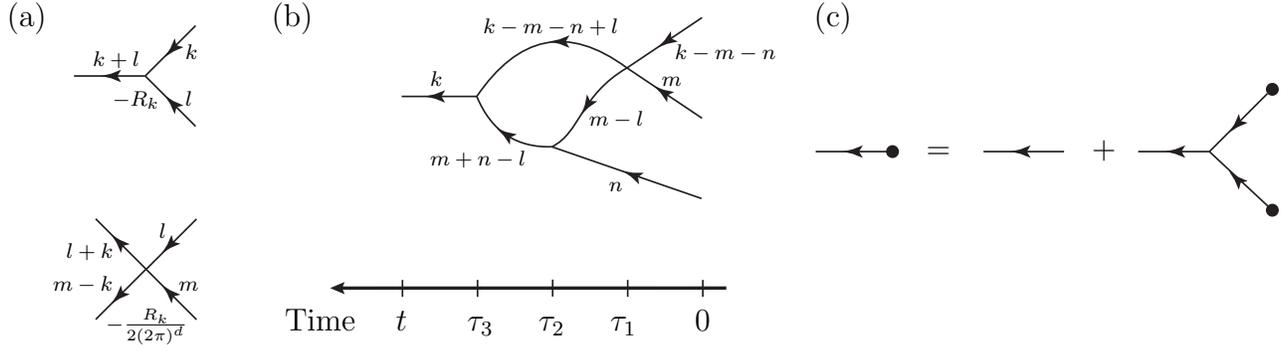

For the model $A+A\rightarrow\phi$ we are considering, perturbative expansion is represented by Feynman diagrams, as indicated in Fig. \ref{PertFig}, where we see the basic nodes, a typical diagram and a recursion relation for all loop free diagrams. 

In Fig. \ref{PertFig}(a) we have nodes of degrees three and four corresponding to the cubic and quartic terms $\Omega_R$ and $\Xi_R$. Note that in the cubic term factor $R_k$ utilises one of the incoming momenta, whereas in the quartic term, the factor $R_k$ introduces a momenta to be integrated over.

In Fig. \ref{PertFig}(b) we have an example of a Feynman diagram of third order contributing to the density. Note that the total momenta in any time interval is conserved (and equal to the final momentum $k$ for the density $X(k;t)$). With three incoming edges and a loop this gives three momenta to be integrated over. Being a third order term means we have four time intervals. This results in the following contribution to density,
\begin{equation}
\frac{-1}{2(2\pi)^d}\iint_{\Upsilon^3}\dd l \hh \dd m \hh \dd n \hh R_l R_mR_n T(k,l,m,n;t)
v_{k-m-n}v_mv_n,
\nonumber
\end{equation}
where the time term $T(k,l,m,n;t)$, which contains the product of propagators, can be written as,
\begin{eqnarray}
&& \iint_{t>\tau_3>\tau_2>\tau_1>0}\dd\tau_3\hh\dd\tau_2\hh\dd\tau_1\hh
e^{-(t-\tau_3)Dk^2}
e^{-(\tau_3-\tau_2)D[(k-m-n+l)^2+(m+n-l)^2]}\cdot
\nonumber\\
&& \hspace{5cm} 
e^{-(\tau_2-\tau_1)D[(k-m-n+l)^2+(m-l)^2+n^2]}
e^{-\tau_1D[(k-m-n)^2+m^2+n^2]}
\nonumber\\
& = & e^{-tDk^2}*e^{-tD[(k-m-n+l)^2+(m+n-l)^2]}
*e^{-tD[(k-m-n+l)^2+(m-l)^2+n^2]}*e^{-tD[(k-m-n)^2+m^2+n^2]}
\nonumber\\
& = & L^{-1}\left\{ 
L(e^{-tDk^2}) L(e^{-tD[(k-m-n+l)^2+(m+n-l)^2]})\right.\cdot
\nonumber\\
&& \hspace{2cm}
\left.L(e^{-tD[(k-m-n+l)^2+(m-l)^2+n^2]})L(e^{-tD[(k-m-n)^2+m^2+n^2]})
\right\}.
\nonumber
\end{eqnarray}
Note that each exponent contains the total squared momenta present in the Feynman diagram for that time period. For example (in the second exponential), $(k-m-n+l)^2+(m+n-l)^2$ is the total squared momenta from the two propagator arms present across the time interval $[\tau_2,\tau_3]$. As the time integral is over a simplex, which can therefore be represented as a convolution product, the factor $T(k,l,m,n;t)$ can subsequently be explicitly derived with the aid of a Laplace transform, partial fractions and Laplace inversion (details are left to the reader). This then leaves integration over momenta, which can be complicated, depending on the nature of the interaction function $R_p$. We do not explore this further, but note that there are two situations where things will simplify, one is the diffusion limiting case where $R_p$ is a delta function, and the other is where the initial density $v$ is uniform, which has the effect that external momenta on the right of diagrams are zero. 

In Fig. \ref{PertFig}(c) we have the standard Dyson recursion representing all loop free diagrams (see \cite{Cardy2006}, for example). This results in an equation of the form,
\begin{equation}
X(k;t) = e^{-Dtk^2}v_k + \int_0^t \dd s \hh e^{-D(t-\tau)k^2}e^{-D(\tau-s)[(k-m)^2+m^2]}R_mX(m,s)X(k-m,s).
\nonumber
\end{equation}
If this is differentiated with respect to time and Fourier inverted back to positional space, the following equation is obtained,
\begin{equation}
\frac{\partial X(p;t)}{\partial t} = D\Delta X(p;t) + \int_0^t \dd s \hh 
\int_\Upsilon \dd q X(q;s)\Phi_D(p-q;t-s)
\int_\Upsilon \dd z R_{p-z}\int_\Upsilon X(r;s)\Phi_D(z-r;t-s),
\nonumber
\end{equation} 
where $\Phi_D(x,t) = (4\pi Dt)^{-d/2}\exp(\frac{-x^2}{4Dt})$ is the heat kernel. Note in the last term we have a particle at position $q$ at time $s$ diffusing to position $p$, and a particle at position $r$ diffusing to position $z$, along with an interaction between positions $p$ and $z$, integrated over possible times $s$. Thus we find that the open diagrams produce a mean field equation, as observed elsewhere \cite{Cardy2006, Doi1976b}.

%%%%%%%%%%%%%%%%%%%%%%%%%%%%%%%%%%%%%%%%%%%%%%%%%%%%%%%%%%%%%
%%%%%%%%%%%%%%%%%%%%%%%%%%%%%%%%%%%%%%%%%%%%%%%%%%%%%%%%%%%%%

\section{Conclusions}

This work reports a field theoretic approach to reaction diffusion systems. Specifically, the array of operators seen in Doi field theory can be reinterpreted with the It\^{o} calculus machinery of quantum stochastic processes. This involves extending the usual three classes of noise associated with quantum stochastic processes, which are not necessarily associative or finite in number for certain systems. However, this results in techniques applicable to an array of systems.

For local processes, where interactions occur at a single location, a product expansion method provides a powerful alternative to path integral techniques. For example, it provides a convenient method to describe spatially varying birth-death models with time dependent rates, which are difficult to model by other methods. Although non-spatial time dependent models can be approached with Wei-Norman methods \cite{Ohkubo2014}, extracting exact results for spatial dependent models would then likely require perturbative diagram summation tricks, or possibly exact methods of path integration. For discrete processes without spatial dependency, the machinery also provides utility, producing algebraic reduction techniques which removes the need for path integration. 

For non-local processes, where interactions occur between physically separated particles, the product expansion methods are not tractable. However, perturbation theory is still possible with the use of quantum It\^{o} calculus. In particular, the interaction picture and time series expansion methods of Dyson can be formulated this way, as seen with the model $A+A\rightarrow \phi$ used in Doi \cite{Doi1976b}. Notably, the Doi shift was not implemented in Doi's original formulation, and implementing this with perturbation analysis was seen to significantly simplify the resultant expansions.

Real coherent states were utilised in this exposition. For methods without path integrals (such as those reported here, or Doi \cite{Doi1976a, Doi1976b}), they provide sufficient flexibility. However, complex coherent states can be useful. For example, when constructing path integrals, resolutions of identity can involve coherent state integration over the entire complex domain. Note that the machinery developed here is applicable in full complex form (such as that stated in Eq. (\ref{CohNorm})), which may prove useful in future work.

For many stochastic quantum processes, the space $\Upsilon$ is of dimension one, usually being treated as a time variable, meaning systems can be reformulated as quantum stochastic differential equations. How to find such an interpretation for higher dimensional spaces is unclear. Conversely, the processes involved in this work are notably simpler than standard quantum It\^{o} methods in one regard; they do not employ an initial space, which has proved fruitful in many areas of application. Evans-Hudson flows, for example, have been used to describe non-spatial birth death processes, suggesting the It\^{o} machinery developed above may be adaptable to more complex reaction diffusion systems of interest.

Reaction diffusion systems can be tackled with a range of other methods, including master equations, path integral techniques, perturbative techniques, semi-group methods, system size expansion methods, and numerical techniques, to name a few. However, the It\^{o} calculus methods described in this work clearly provide a useful addition to this bag of tricks.

%%%%%%%%%%%%%%%%%%%%%%%%%%%%%%%%%%%%%%%%%%%%%%%%%%%%%%%%%%%%%
%%%%%%%%%%%%%%%%%%%%%%%%%%%%%%%%%%%%%%%%%%%%%%%%%%%%%%%%%%%%%

\appendix
\section*{Appendix: Path Integral for Death-Diffusion Process}

The death-diffusion process $A \rightarrow \phi$ with constant death rate $\mu$ and diffusion rate $D$ has Liouvillian $\mathcal{L} = \int_\Upsilon \left(\mu(a_p-a_p^\dag a_p)+D a_p^\dag\Delta a_p\right)$. The generating functional for evolution operator $U = e^{t\mathcal{L}}$ can be written as $\braket{y|U|w}$. To evaluate this via path integration, the following resolution of identity is used \cite{Greenman2017}, where $\Upsilon \equiv \mathbb{R}^d$,
\begin{equation}
I = \iint_{\Upsilon^2} \mathcal{D} u \hh \mathcal{D} v \hh e^{-i\int_\chi uv \hh \ddd p}\ket{iv}\bra{u}.
\nonumber
\end{equation}

Then standard time slicing results in the following expression (with $u_{N+1}\equiv y$ and $N \hh \dd t = t$),
\begin{equation}
\braket{y|U|w} = \prod_{i=0}^N \iint_{\Upsilon^2} \mathcal{D} u_i \hh \mathcal{D} v_i \hh e^{-i\int_\Upsilon u_i(p)v_i(p)\hh\ddd p}\braket{u_{i+1}|e^{\ddd t\hh\mathcal{L}}|iv_i} \braket{u_0|w}.
\nonumber
\end{equation}
Then from the standard operation of creation and annihilation operators on coherent states, this becomes, in the formal limit of $N \rightarrow \infty$,
\begin{equation}
\braket{y|U|w} = \iint \mathcal{D}u \mathcal{D}v \exp\left\{ i\int_0^t \dd s \hh \int_\Upsilon \dd p \hh v\left[\mu(1-u)+D\Delta u + \frac{\partial u}{\partial s}\right] + \int_\Upsilon \hh \dd p \hh u_0w\right\}.
\nonumber
\end{equation}
Note that there is an integration by parts to get the third term, and two integration by parts to move the Laplacian action from the $v$ to the $u$ function.

Then if we do the path integration with respect to the $v$ variable, we get a delta functional $\delta(u-u')$ such that,
\begin{equation}
\begin{cases}
\displaystyle\frac{\partial u'}{\partial s} = -D\Delta u' - \mu(1-u'), \\
u'(p;t) = y(p).
\end{cases}
\label{DeltaCond}
\end{equation}
Thus we have a reverse time heat equation with a source. After a bit of manipulation, we get solution,
\begin{equation}
u'(p,s) = e^{-\mu(t-s)}\int_{\Upsilon}\Phi(p-q;t-s)y(q) \hh \dd q + 1-e^{-\mu(t-s)},
\nonumber
\end{equation}
where $\Phi(p;s) = \frac{1}{(4\pi D s)^{d/2}}\exp(\frac{-|p|^2}{4D s})$ is the standard heat equation solution for a point source. Finally, substituting for $u_0 \equiv u'(p;0)$ (that is, doing the $u$ path integration to substitute the condition in Eq. (\ref{DeltaCond}) forced by the delta functional), results in,
\begin{equation}
\braket{y|U|w}e^{-\int_\Upsilon w_p \ddd p} = \exp\left\{ e^{-\mu t}\iint_{\Upsilon^2} w(p) \Phi(p-q;t) y(q) \hh \dd p \hh \dd q - e^{-\mu t}\int_\Upsilon w \hh \dd p \right\},
\nonumber
\end{equation}
giving agreement with the expression in Eq. (\ref{DeathGenFn}).

%%%%%%%%%%%%%%%%%%%%%%%%%%%%%%%%%%%%%%%%%%%%%%%%%%%%%%%%%%%%%
%%%%%%%%%%%%%%%%%%%%%%%%%%%%%%%%%%%%%%%%%%%%%%%%%%%%%%%%%%%%%

{\footnotesize
	\bibliographystyle{acm}
	\bibliography{refs_DoiQSP}
}

\end{document}